\documentclass[useAMS,usenatbib]{mn2e}
\usepackage{graphicx}

\title[Precice mass and radius values for both components of the pre-CV NN Ser]{Precise mass and radius values for the white dwarf and low mass M dwarf in the pre-cataclysmic binary NN Serpentis}

\author[S. G. Parsons et al.]{S.~G.~Parsons$^{1}$\thanks{steven.parsons@warwick.ac.uk},
T.~R.~Marsh$^{1}$,
C.~M.~Copperwheat$^{1}$,
V.~S.~Dhillon$^{2}$,
\newauthor
S.~P.~Littlefair$^{2}$,
B.~T.~G{\"a}nsicke$^{1}$
and R.~Hickman$^{1}$ \\
$^{1}$Department of Physics, University of Warwick, Coventry, CV4 7AL \\
$^{2}$Department of Physics and Astronomy, University of Sheffield, Sheffield S3 7RH}
\begin{document}
\input{pjw_aas_macros.cls}
\date{Accepted 2009 November 18. Received 2009 November 17; in original form 2009 September 22}

\pagerange{\pageref{firstpage}--\pageref{lastpage}} \pubyear{2009}

\maketitle

\label{firstpage}

\begin{abstract}
Using the high resolution Ultraviolet and Visual Echelle Spectrograph (UVES) mounted on the Very Large Telescope (VLT) in combination with photometry from the high-speed CCD camera ULTRACAM, we derive precise system parameters for the pre-cataclysmic binary, NN Ser. A model fit to the ULTRACAM light curves gives the orbital inclination as i$ = 89.6^{\circ} \pm 0.2^{\circ}$ and the scaled radii, $R_\mathrm{WD}/a$ and $R_\mathrm{sec}/a$. Analysis of the HeII 4686\AA~ absorption line gives a radial velocity amplitude for the white dwarf of $K_\mathrm{WD}= 62.3 \pm 1.9$ km$\,$s$^{-1}$. We find that the irradiation-induced emission lines from the surface of the secondary star give a range of observed radial velocity amplitudes due to differences in optical depths in the lines. We correct these values to the centre of mass of the secondary star by computing line profiles from the irradiated face of the secondary star. We determine a radial velocity of $K_\mathrm{sec}= 301 \pm 3$ km$\,$s$^{-1}$, with an error dominated by the systematic effects of the model. This leads to a binary separation of a $= 0.934 \pm 0.009$ R$_{\sun}$, radii of R$_\mathrm{WD} = 0.0211 \pm 0.0002$ R$_{\sun}$ and R$_\mathrm{sec} = 0.149 \pm 0.002$ R$_{\sun}$ and masses of M$_\mathrm{WD} = 0.535 \pm 0.012$ M$_{\sun}$ and M$_\mathrm{sec} = 0.111 \pm 0.004$ M$_{\sun}$. The masses and radii of both components of NN Ser were measured independently of any mass-radius relation. For the white dwarf, the measured mass, radius and temperature show excellent agreement with a `thick' hydrogen layer of fractional mass M$_\mathrm{H}/\mathrm{M}_\mathrm{WD} = 10^{-4}$. The measured radius of the secondary star is 10\% larger than predicted by models, however, correcting for irradiation accounts for most of this inconsistency, hence the secondary star in NN Ser is one of the first precisely measured very low mass objects (M $\la 0.3$ M$_{\sun}$) to show good agreement with models. ULTRACAM \emph{r'}, \emph{i'} and \emph{z'} photometry taken during the primary eclipse determines the colours of the secondary star as (\emph{r'-i'})$_\mathrm{sec} = 1.4 \pm 0.1$ and (\emph{i'-z'})$_\mathrm{sec} = 0.8 \pm 0.1$ which corresponds to a spectral type of M$4\pm0.5$. This is consistent with the derived mass, demonstrating that there is no detectable heating of the unirradiated face, despite intercepting radiative energy from the white dwarf which exceeds its own luminosity by over a factor of 20.
\end{abstract}


\section{Introduction}

Precise measurements of masses and radii are of fundamental importance to the theory of stellar structure and evolution. Mass-radius relations are routinely used to estimate the masses and radii of stars and stellar remnants, such as white dwarfs. Additionally, the mass-radius relation for white dwarfs has played an important role in estimating the distance to globular clusters \citep{Renzini96} and the determination of the age of the galactic disk \citep{Wood92}. However, the empirical basis for this relation is uncertain \citep{Schmidt96} as there are very few circumstances where both the mass and radius of a white dwarf can be measured independently and with precision.

\citet{Provencal98} used \emph{Hipparcos} parallaxes to determine the radii for 10 white dwarfs in visual binaries or common proper-motion (CPM) systems and 11 field white dwarfs. They were able to improve the radii measurements for the white dwarfs in visual binaries and CPM systems. However, they remarked that mass determinations for field white dwarfs are indirect, relying on complex model atmosphere predictions. They were able to support the mass-radius relation on observational grounds more firmly, though they explain that parallax remains a dominant source of uncertainty, particularly for CPM systems. Improvements in our knowledge of the masses and radii of white dwarfs requires additional measurements. One situation where this is possible is in close binary systems. In these cases, masses can be determined from the orbital parameters and radii from light-curve analysis. Of particular usefulness in this regard are eclipsing post-common envelope binaries (PCEBs). The binary nature of these objects helps determine accurate parameters and, since they are detached, they lack the complications associated with interacting systems such as cataclysmic variables. The inclination of eclipsing systems can be constrained much more strongly than for non-eclipsing systems.  Furthermore, the distance to the system does not have to be known, removing the uncertainty due to parallax.

An additional benefit of studying PCEBs is that under favourable circumstances, not only are the white dwarf's mass and radius determined independently of any model, so too are the mass and radius of its companion. These are often low mass late-type stars, for which there are few precise mass and radius measurements. There is disagreement between models and observations of low mass stars; the models tend to under predict the radii by as much as 20-30\% \citep{Lopez07}. Hence detailed studies of PCEBs can lead to improved statistics for both white dwarfs and low mass stars. Furthermore, models of low mass stars are important for understanding the late evolution of mass transferring binaries such as cataclysmic variables \citep{Littlefair08}.

The PCEB NN Ser (PG 1550+131) is a low mass binary system consisting of a hot white dwarf primary and a cool M dwarf secondary. It was discovered in the Palomar Green Survey \citep{Green82} and first studied in detail by \citet{Haefner89} who presented an optical light curve showing the appearance of a strong heating effect and very deep eclipses ($>$4.8mag at $\lambda$ $\sim$ 6500 \AA ). Haefner identified the system as a pre-cataclysmic binary with an orbital period of 0.13d. The system parameters were first derived by \citet{Wood91} using low-resolution ultra-violet spectra then refined by \citet{Catalan94} using higher resolution optical spectroscopy. \citet{Haefner04} further constrained the system parameters using the FORS instrument at the Very Large Telescope (VLT) in combination with high-speed photometry and phase-resolved spectroscopy. However, they did not detect the secondary eclipse leading them to underestimate the binary inclination and hence overestimate the radius and ultimately the mass of the secondary star. They were also unable to directly measure the radial velocity amplitude of the white dwarf and were forced to rely upon a mass-radius relation for the secondary star. Recently, \citet{Brinkworth06} performed high time resolution photometry of NN Ser using the high-speed CCD camera ULTRACAM mounted on the William Herschel Telescope (WHT). They detected the secondary eclipse leading to a better constraint on the inclination, and also detected a decrease in the orbital period which they determined was due either to the presence of a third body, or to a genuine angular momentum loss. Since NN Ser belongs to the group of PCEBs which is representative for the progenitors of the current cataclysmic variable (CV) population \citep{Schreiber03}, the system parameters are important from both an evolutionary point of view as well as providing independent measurements of the masses and radii of the system components.

In this paper we present high resolution VLT/UVES spectra and high time resolution ULTRACAM photometry of NN Ser. We use these to determine the system parameters directly and independently of any mass-radius relations. We compare our results with models of white dwarfs and low mass stars.

\section{Observations and their Reduction}

\subsection{Spectroscopy}

\begin{table}
 \centering
  \caption{Journal of VLT/UVES spectroscopic observations.}
  \label{spec_obs}
  \begin{tabular}{@{}lcccl@{}}
  \hline
  Date&Start&End &No. of &Conditions \\
      &(UT) &(UT)&spectra&(Transparency, seeing) \\
 \hline
 30/04/2004&05:15&08:30&40&Fair, $\sim$2.5 arcsec \\
 01/05/2004&03:14&06:31&40&Variable, 1.2-3.1 arcsec \\
 17/05/2004&03:37&06:58&40&Fair, $\sim$2.1 arcsec \\
 26/05/2004&00:52&04:14&40&Fair, $\sim$2.2 arcsec\\
 27/05/2004&03:14&06:42&40&Good, $\sim$1.1 arcsec \\
 10/06/2004&02:06&05:22&40&Good, $\sim$1.1 arcsec \\
 12/06/2004&02:39&05:54&40&Fair, $\sim$2.1 arcsec \\
 15/06/2004&02:41&05:56&40&Good, $\sim$1.8 arcsec \\
 27/06/2004&03:54&05:11&15&Fair, $\sim$2.2 arcsec \\
\hline
\end{tabular}
\end{table}

Spectra were taken in service mode over nine different nights between 2004 April and June using the Ultraviolet and Visual Echelle Spectrograph (UVES) installed at the European Southern Observatory Very Large Telescope (ESO VLT) 8.2-m telescope unit on Cerro Paranal in Chile \citep{Dekker00}. In total 335 spectra were taken in each arm, details of these observations are listed in Table~\ref{spec_obs}. Observation times were chosen to cover a large portion of the orbital cycle and an eclipse was recorded on each night. Taken together the observations cover the whole orbit of NN Ser. Exposure times of 250.0s and 240.0s were used for the blue and red spectra respectively; these were chosen as a compromise between orbital smearing and signal-to-noise ratio (\emph{S/N}). The wavelength range covered is 3760--4990\AA~ in the blue arm and 6710--8530 and 8670--10400\AA~ in the red arm. The reduction of the raw frames was conducted using the most recent release of the UVES Common Pipeline Library (CPL) recipes (version 4.1.0) within ESORex, the ESO Recipe Execution Tool, version 3.6.8. The standard recipes were used to optimally extract each spectrum. A ThAr arc lamp spectrum was used to wavelength calibrate the spectra. Master response curves were used to correct for detector response and initially flux calibrate the spectra since no standard was observed. The spectra have a resolution of R$\sim$80,000 in the blue and R$\sim$110,000 in the red. Orbital smearing limits the maximum resolution; at conjunction lines will move by at most $\sim37$ km$\,$s$^{-1}$. Since the widths of the lines seen in the spectra are at least $\sim100$ km$\,$s$^{-1}$, this effect is not large. The \emph{S/N} becomes progressively worse at longer wavelengths and a large region of the upper red CCD spectra was ignored since there was very little signal. The orbital phase of each spectrum was calculated using the ephemeris of \citet{Brinkworth06}. 

The spectral features seen are similar to those reported by \citet{Catalan94} and \citet{Haefner04}: Balmer lines, which appear as either emission or absorption depending upon the phase, HeI and CaII emission lines and HeII 4686\AA~ in absorption. The Paschen series is also seen in emission in the far red. In addition, MgII 4481\AA~ emission is seen as well as a number of fainter MgII emission lines beyond 7800\AA~. Weak FeI emission lines are seen throughout the spectrum and faint CI emission is seen beyond 8300\AA~ (see Table~\ref{lines} for a full list of identified emission lines). The strength of all the emission lines is phase-dependent, peaking at phase 0.5, when the heated face of the secondary star is in full view, then disappearing around the primary eclipse. Several sharp absorption features are observed not to move over the orbital period, these are interstellar absorption features and include interstellar CaII absorption.

\subsection{Blaze Removal}

\begin{figure}
  \begin{center}
    \includegraphics[width=\columnwidth]{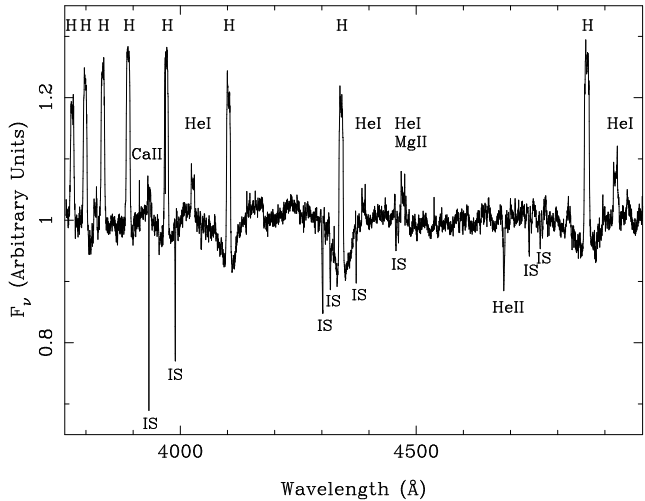} \\
  \caption{Averaged, normalised UVES blue arm spectrum with the blaze removed. IS corresponds to interstellar absorption features. The discontinuity at $\sim 4150$\AA~ and the emission-like feature at $\sim 4820$\AA~ are most likely instrumental features or artifacts of the UVES reduction pipeline as they are seen in all 335 spectra.}
  \label{blaze}
  \end{center}
\end{figure}

An echelle grating produces a spectrum that drops as one moves away from the blaze peak, this is known as the blaze function. After reduction a  residual ripple pattern was visible in the blue spectra corresponding to the blaze function. This was approximately removed by fitting with a sinusoid of the form
\begin{eqnarray}
B(\lambda)& = &a_{0} + a_{1}\sin(2\pi\phi) + a_{2}\lambda\sin(2\pi\phi) \\
& & +\; a_{3}\cos(2\pi\phi) + a_{4}\lambda\cos(2\pi\phi). \nonumber
\end{eqnarray}
The phase ($\phi$) was calculated by identifying the central wavelength of each echelle order. The line table produced using the ESORex recipe uves\_cal\_wavecal provided this information. Then using the relation
\begin{eqnarray}
\lambda_{n}(O-n) = c,
\end{eqnarray}
where $c$ and $O$ are constants and $\lambda_{n}$ is the central wavelength of order $n$, gives us the phase. We find values of $O = 125$ and $c = 465700$, which are similar for all the spectra. Therefore the phase of the ripple is
\[\phi = 125 - \frac{465700}{\lambda}.\]
Since the phase is now known, Equation 1 reduces to a simple linear fit. Figure~\ref{blaze} is a normalised average of all the UVES blue arm spectra with the blaze removed. Since this is only a simple fit some residual pattern does remain after division by the blaze function but overall the effect is greatly reduced.

\subsection{Photometry}

\begin{table*}
 \centering
  \caption{ULTRACAM observations of NN Ser. The primary eclipse occurs at phase 1, 2 etc.}
  \label{pho_obs}
  \begin{tabular}{@{}lccccccl@{}}
  \hline
  Date&Filters&Telescope&UT     &UT   &Average     &Phase &Conditions \\
      &       &         &start  &end  &exp time (s)&range &(Transparency, seeing)           \\
 \hline
 17/05/2002 & \emph{u'g'r'} & WHT & 21:54:40 & 02:07:54 & 2.4 & 0.85--2.13 & Good, $\sim$1.2 arcsec \\
 18/05/2002 & \emph{u'g'r'} & WHT & 21:21:20 & 02:13:17 & 3.9 & 0.39--1.23 & Variable, 1.2-2.4 arcsec \\
 19/05/2002 & \emph{u'g'r'} & WHT & 23:58:22 & 00:50:52 & 2.0 & 0.93--1.10 & Fair, $\sim$2 arcsec \\
 20/05/2002 & \emph{u'g'r'} & WHT & 00:58:23 & 01:57:18 & 2.3 & 0.86--1.14 & Fair, $\sim$2 arcsec \\
 19/05/2003 & \emph{u'g'z'} & WHT & 22:25:33 & 01:02:25 & 6.7 & 0.47--1.12 & Variable, 1.5-3 arcsec \\
 21/05/2003 & \emph{u'g'i'} & WHT & 00:29:00 & 04:27:32 & 1.9 & 0.32--0.59 & Excellent, $\sim$1 arcsec \\
 22/05/2003 & \emph{u'g'i'} & WHT & 03:24:57 & 03:50:40 & 2.0 & 0.36--0.08 & Excellent, $<$1 arcsec \\
 24/05/2003 & \emph{u'g'i'} & WHT & 22:58:55 & 23:33:49 & 2.0 & 0.90--0.08 & Good, $\sim$1.2 arcsec \\
 25/05/2003 & \emph{u'g'i'} & WHT & 01:29:45 & 02:15:58 & 2.0 & 0.39--0.64 & Excellent, $\sim$1 arcsec \\
 03/05/2004 & \emph{u'g'i'} & WHT & 22:13:44 & 05:43:11 & 2.5 & 0.37--2.27 & Variable, 1.2-3.2 arcsec \\
 04/05/2004 & \emph{u'g'i'} & WHT & 23:18:46 & 23:56:59 & 2.5 & 0.91--0.61 & Variable, 1.2-3 arcsec \\
 09/03/2006 & \emph{u'g'r'} & WHT & 01:02:34 & 06:46:49 & 2.0 & 0.91--2.70 & Variable, 1.2-3 arcsec \\
 10/03/2006 & \emph{u'g'r'} & WHT & 05:01:13 & 05:50:14 & 2.0 & 0.85--1.11 & Excellent, $<$1 arcsec \\
 09/06/2007 & \emph{u'g'i'} & VLT & 04:59:25 & 05:46:18 & 0.9 & 0.40--0.61 & Excellent, $\sim$1 arcsec \\
 16/06/2007 & \emph{u'g'i'} & VLT & 03:57:48 & 04:54:39 & 2.0 & 0.86--1.15 & Good, $\sim$1.2 arcsec \\
 17/06/2007 & \emph{u'g'i'} & VLT & 01:50:16 & 02:38:09 & 1.0 & 0.86--1.11 & Excellent, $<$1 arcsec \\
 07/08/2008 & \emph{u'g'r'} & WHT & 23:41:29 & 00:22:46 & 2.8 & 0.87--1.07 & Excellent, $<$1 arcsec \\
\hline
\end{tabular}
\end{table*}

The data presented here were collected with the high speed CCD camera ULTRACAM \citep{Dhillon07}, mounted as a visitor instrument on the $4.2$m William Herschel Telescope (WHT) and on the VLT in June 2007. A total of ten observations were made in 2002 and 2003, and these data were supplemented with observations made at a rate of $\sim 1$ -- $2$ a year up until 2008. ULTRACAM is a triple beam camera and most of our observations were taken simultaneously through the SDSS \emph{u'}, \emph{g'} and \emph{i'} filters. In a number of instances an \emph{r'} filter was used in place of \emph{i'}; this was mainly for scheduling reasons. Additionally, a \emph{z'} filter was used in place of \emph{i'} for one night in 2003.

A complete log of the observations is given in Table~\ref{pho_obs}. We windowed the CCD in order to achieve an exposure time of $2$ -- $3$s, which we varied to account for the conditions. The dead time was $\sim 25$ms.

All of these data were reduced using the ULTRACAM pipeline software. Debiassing, flatfielding and sky background subtraction were performed in the standard way. The source flux was determined with aperture photometry using a variable aperture, whereby the radius of the aperture is scaled according to the FWHM. Variations in observing conditions were accounted for by determining the flux relative to a comparison star in the field of view. There were a number of additional stars in the field which we used to check the stability of our comparison. For the flux calibration we determined atmospheric extinction coefficients in the \emph{u'}, \emph{g'} and \emph{r'} bands and subsequently determined the absolute flux of our targets using observations of standard stars (from \citealt{Smith02}) taken in twilight. We use this calibration for our determinations of the apparent magnitudes of the two sources, although we present all light curves in flux units determined using the conversion given in \citet{Smith02}. Using our absorption coefficients we extrapolate all fluxes to an airmass of $0$. The systematic error introduced by our flux calibration is  $<0.1$ mag in all bands. For all data we convert the MJD times to the barycentric dynamical timescale, corrected to the solar system barycentre.

\begin{figure}
\begin{center}
 \includegraphics[width=\columnwidth]{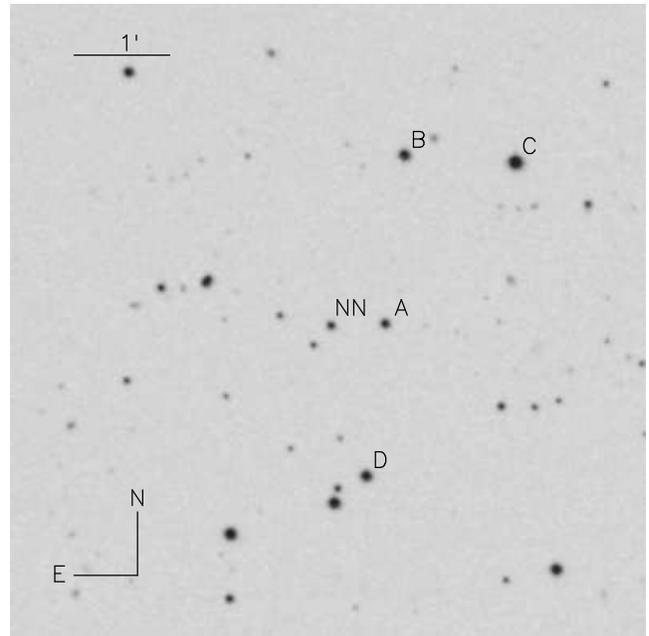}
 \caption{Digital Sky Survey finding chart (POSS II, \emph{blue}) for NN Ser. Comparison stars are marked.}
 \label{finding}
 \end{center}
\end{figure}

A number of comparison stars were observed, their locations are shown in Figure~\ref{finding} and details of these stars are given in Table~\ref{cstars}. Where possible we use comparison star C since it is brighter. However, in 2002 only comparison star D was observed and in the 2007 VLT data, comparison stars C and B were saturated in \emph{g'} and \emph{i'}. We therefore use star C for the comparison in the \emph{u'} and star A for the \emph{g'} and \emph{i'}.

The light curves were corrected for extinction differences by using the comparison star observations. A first-order polynomial was fit to the comparison star photometry in order to determine the comparison star's colours (magnitudes listed in Table~\ref{cstars}). The colour of the white dwarf in NN Ser was calculated by fitting a zeroth-order polynomial to the flat regions around the primary eclipse with a correction made in the \emph{r'} and \emph{i'} bands for the secondary stars contribution (the contribution of the secondary star in the \emph{u'} and \emph{g'} bands around the primary eclipse is negligible). The colour dependant difference in extinction coefficients for the comparison star and NN Ser were then calculated from a theoretical extinction vs. colour plot\footnote{theoretical extinction vs. colour plots for ULTRACAM are available at http://garagos.net/dev/ultracam/filters}. The additional extinction correction for NN Ser for each night is then
\begin{eqnarray}
10^{\frac{1}{2.5}\left(k_{N} - k_{C}\right)X},
\end{eqnarray}
where k$_{N}$ is the extinction coefficient for NN Ser, k$_{C}$ is the extinction coefficient for the comparison and X is the airmass. The value of $k_{N} - k_{C}$ for each band was similar for all the comparisons used. In the \emph{u'} filter $k_{N} - k_{C} \sim 0.03$, for the \emph{g'} band $k_{N} - k_{C} \sim 0.02$, for the \emph{r'} band $k_{N} - k_{C} \sim 0.002$ and for the \emph{i'} band $k_{N} - k_{C} \sim 0.0005$.

\begin{table}
 \centering
  \caption{Comparison star magnitudes and positional offsets from NN Ser. There were no \emph{i'} band observations of star D. The magnitudes for the white dwarf in NN Ser are shown in Table~\ref{distance}. }
  \label{cstars}
  \begin{tabular}{@{}lcccccc@{}}
  \hline
  Star   & \emph{u'} & \emph{g'} & \emph{r'} & \emph{i'} & RA off. & Dec off. \\
         &           &           &           &           & (arcsec)& (arcsec) \\
  \hline
  A      & 17.0      & 15.6      & 15.8      & 15.0      & -34.1   & +2.2     \\
  B      & 16.0      & 14.7      & 15.1      & 14.3      & -46.4   & +106.7   \\
  C      & 14.6      & 13.4      & 13.7      & 12.8      & -114.5  & +103.7   \\
  D      & 16.7      & 14.6      & 13.7      & --        & -22.2   & -94.1    \\
\hline
\end{tabular}
\end{table}

\begin{figure*}
\begin{center}
 \includegraphics[width=0.95\textwidth]{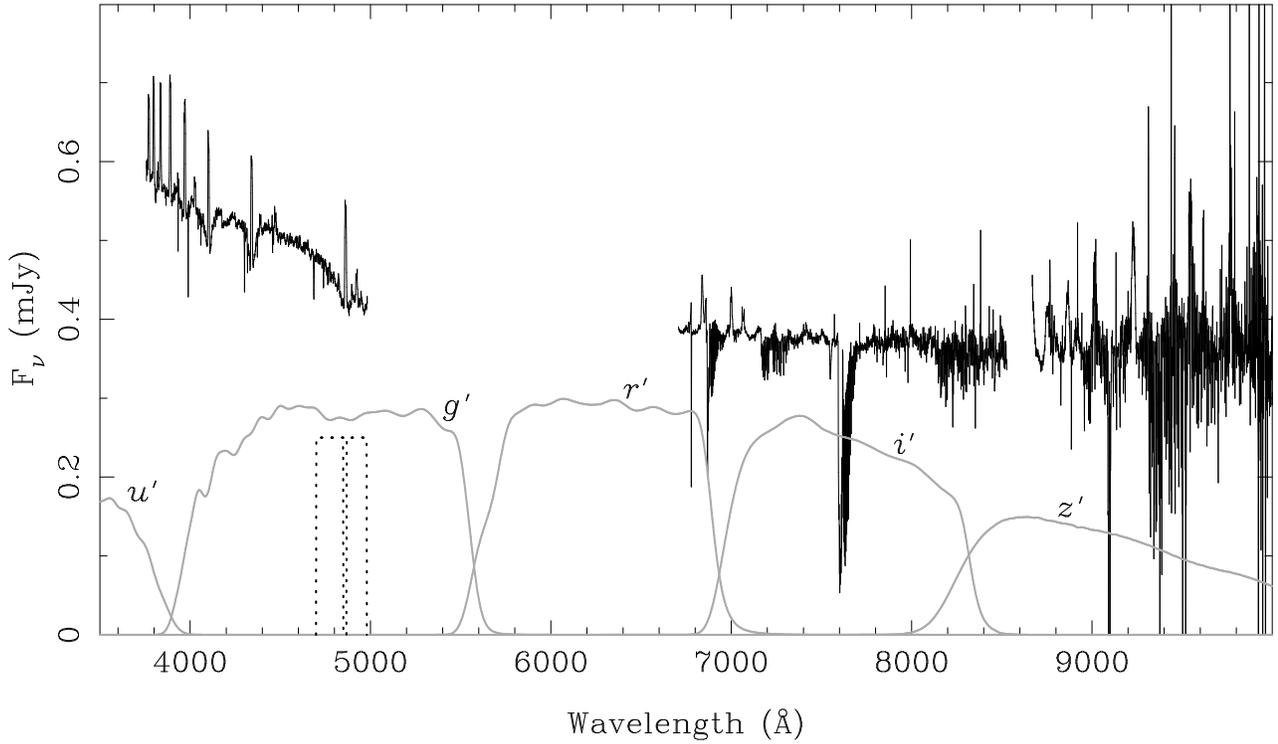}
 \caption{Averaged spectra from the blue, lower and upper red CCD chips with ULTRACAM filter response curves. The spectra were not telluric corrected. The \emph{dotted} line is the filter profile based on the \emph{g'} filter used to flux calibrate the UVES blue CCD spectra since the \emph{g'} filter doesn't cover the same spectra range.}
 \label{avfilt}
 \end{center}
\end{figure*}

\begin{figure}
\begin{center}
 \includegraphics[width=\columnwidth]{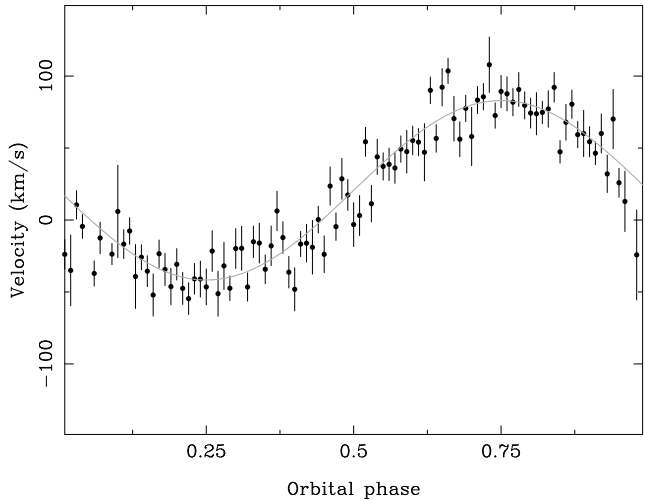}
 \caption{Sine curve fit for the HeII 4686\AA~ absorption line fitted with a straight line and a Gaussian. The measured radial velocity amplitude for the primary is 62.3 $\pm$ 1.9 km$\,$s$^{-1}$. }
 \label{heabs}
\end{center}
\end{figure}

\begin{figure}
\begin{center}
 \includegraphics[width=\columnwidth]{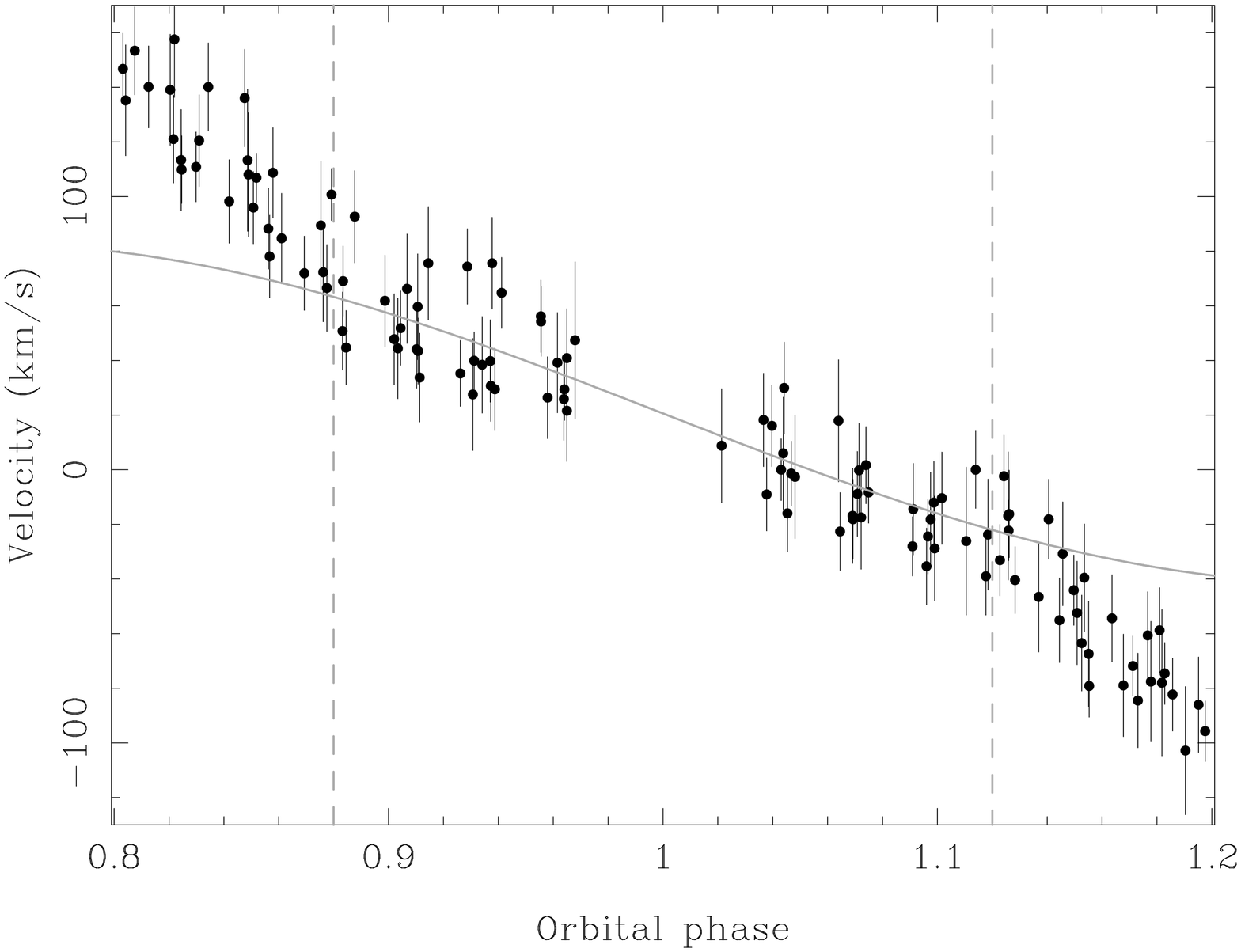}
 \caption{Sine curve fit to the Balmer absorption features near the primary eclipse over-plotted with the radial velocity amplitude of the primary calculated using the HeII line, the deviations away from the eclipse are caused by emission from the secondary star, hence those spectra within the \emph{dashed} lines contain no emission from the secondary star.}
 \label{eclfit}
\end{center}
\end{figure}

The flux-calibrated, extinction-corrected light curves for each filter were phase binned using the ephemeris of \citet{Brinkworth06}. Data within each phase bin were averaged using inverse variance weights whereby data with smaller errors are given larger weightings. Smaller bins were used around both the eclipses. The result of this is a set of high signal-to-noise light curves for NN Ser (one for each filter).

\subsection{Flux Calibration}

The ULTRACAM photometry was used to flux calibrate each of the UVES spectra. Figure~\ref{avfilt} shows the average spectra from each of the detectors (the UVES blue CCD spectra cover 3760--4990\AA~, the lower red CCD spectra cover the range 6710--8530 and the upper red CCD spectra cover 8670--10400\AA~) along with ULTRACAM response curves for each filter. A simple model was fitted to the ULTRACAM \emph{g'} and \emph{i'} light curves (see Section 4.1 for details of the model fitting). The aim of this model was to reproduce the light curve as closely as possible. The model was then used to predict the flux at the times of each of the UVES observations (NN Ser shows no stochastic variations or flaring despite the rapidly rotating secondary star). Since the \emph{i'} filter nicely covers the same spectral range as the UVES spectra from the lower red CCD detector, we could derive synthetic fluxes from the spectra using the standard \emph{i'} filter response. The \emph{g'} filter does not entirely overlap our spectra, therefore we employed a narrower filter profile centred on the middle of the \emph{g'} filter, while avoiding H$\beta$ to prevent its biassing the results. 

\begin{figure*}
\begin{center}
 \includegraphics[width=0.95\textwidth,height=\columnwidth]{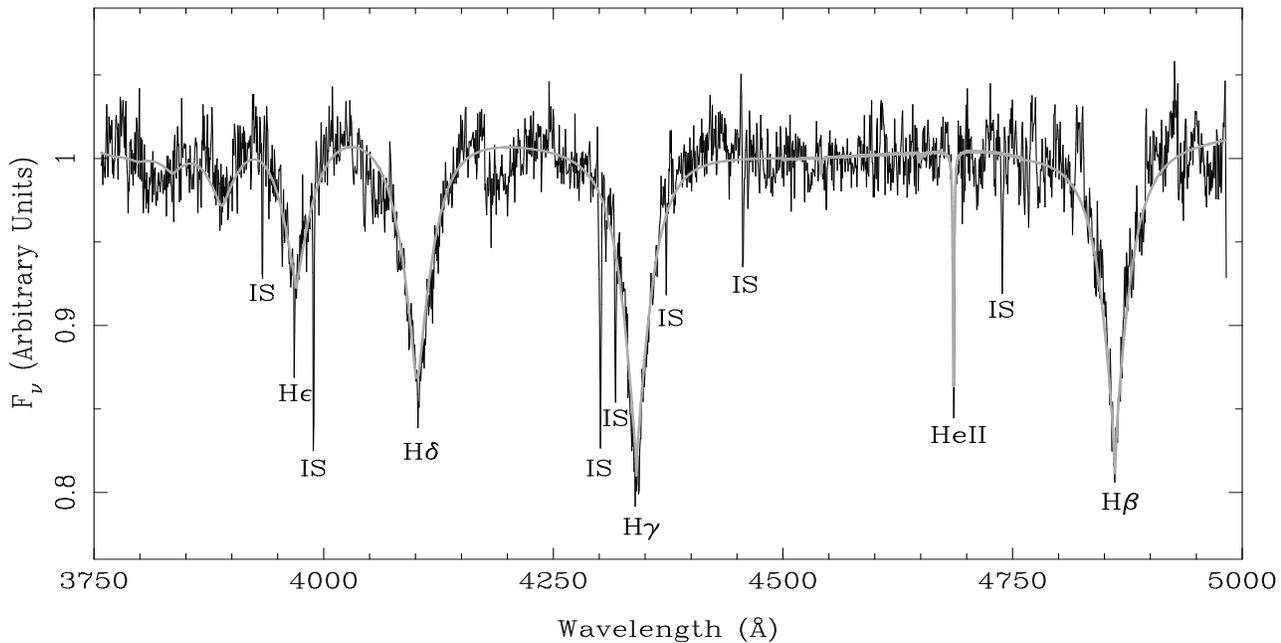}
 \caption{Normalised white dwarf spectrum with an over-plotted $T=57000K$, $\log{g}=7.5$ white dwarf model spectrum including homogeneously mixed helium ($N_\mathrm{He} = 4 \times 10^{-4}$ by number). IS corresponds to interstellar absorption features.}
 \label{wdfull}
\end{center}
\end{figure*}

The average flux of each spectrum was calculated within these filters then compared to the model light curve value for that phase. The spectrum was then multiplied by an appropriate constant to match these values up. Since there is only one \emph{z'} ULTRACAM light curve which, due to conditions, was fairly poor, it was not used to flux calibrate the UVES upper red CCD spectra even though it covers a similar spectra range. Rather, the \emph{i'} model was extrapolated to the longer wavelength range.

\section{Results}

\subsection{The White Dwarf's Spectrum}

In order to recover the white dwarf's spectrum, its radial velocity amplitude must be calculated. Due to the broad nature of the Balmer absorption features and the contamination by emission from the secondary star, we concluded that using the Balmer absorption lines to calculate the radial velocity amplitude of the white dwarf would be likely to give an incorrect result. Fortunately, the narrower HeII 4686\AA~ absorption line is seen in absorption at all phases and no HeII emission is seen at any time throughout the spectra. In addition, there are no nearby features around the HeII 4686\AA~ line making it a good feature to use to calculate the radial velocity amplitude of the white dwarf; it is indeed the only clean feature from the white dwarf that we could identify. We fit the line using a combination of a straight line plus a Gaussian. We use this to calculate a radial velocity amplitude for the white dwarf.

Figure~\ref{heabs} shows the fit to the HeII 4686\AA~ velocities. The radial velocity amplitude found is $62.3 \pm 1.9$ km$\,$s$^{-1}$ significantly smaller than the value of $80.4 \pm 4.1$ km$\,$s$^{-1}$ calculated indirectly by \citet{Haefner04} from their light curve analysis as they relied upon a mass-radius relation for the secondary star. The difference is due to their overestimation of the mass of the secondary star by roughly 30\%. 

Just before and after the primary eclipse the reprocessed light from the secondary star is not yet visible, hence spectra taken at these phases contain just the white dwarf's features. More precise constraints on this range can be made by studying the Balmer lines. The deep wide Balmer absorption lines from the white dwarf are gradually filled in by the emission from the secondary star as the system moves away from the primary eclipse. Multiple Gaussians were fit to the Balmer absorption features, any deviation from this is due to emission from the secondary star which increases the measured velocity amplitude of the lines. Figure~\ref{eclfit} shows this fit along with the HeII 4686\AA~ radial velocity amplitude fit. No deviation is seen before phase $\sim0.12$ and after phase $\sim0.88$ hence spectra taken between these phases contain spectral information on the white dwarf only (except those taken during the eclipse). 

\begin{figure*}
  \begin{center}
   \includegraphics[width=0.9\textwidth]{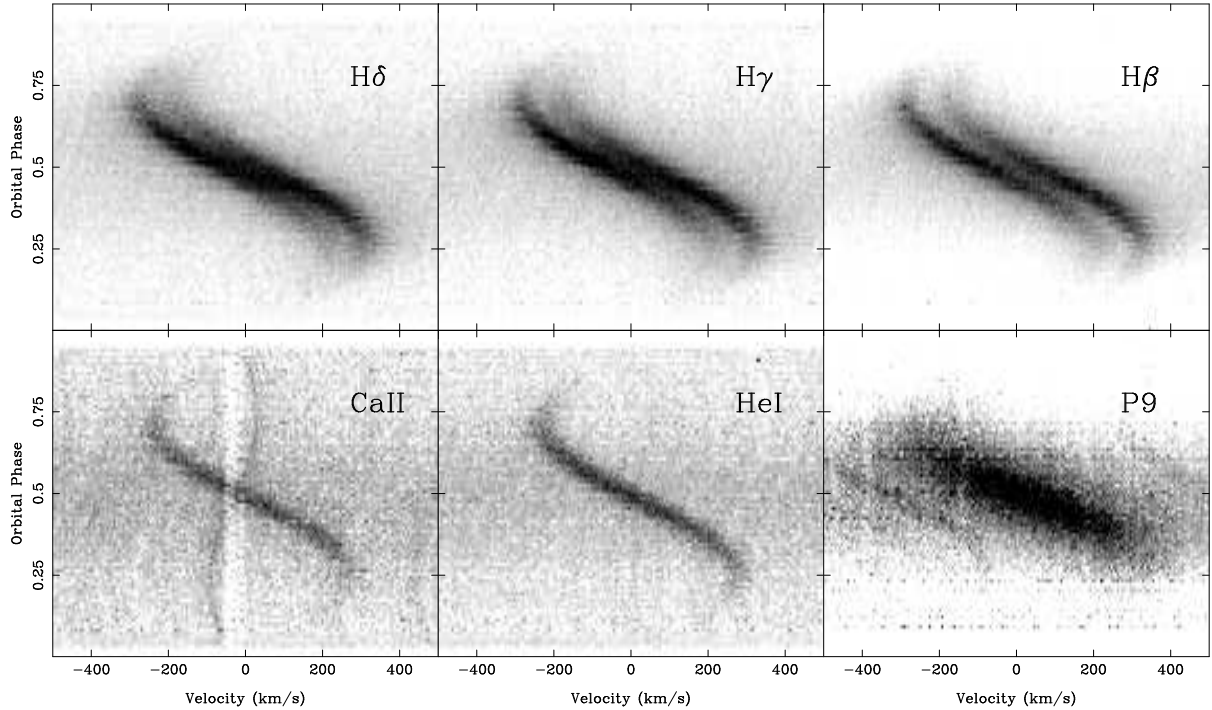}
  \caption{Trailed spectra of various lines. The white dwarf component has been subtracted. White represents a value of 0.0 in all trails. For the Balmer lines and the Paschen line, black represents a value of 2.0, for the other lines, black represents a value of 1.0. The subtraction of the white dwarf component creates a peak on the CaII trail due to the presence of interstellar absorption.}
  \label{trail}
  \end{center}
\end{figure*}

\begin{figure*}
  \begin{center}
    \includegraphics[width=0.327\textwidth]{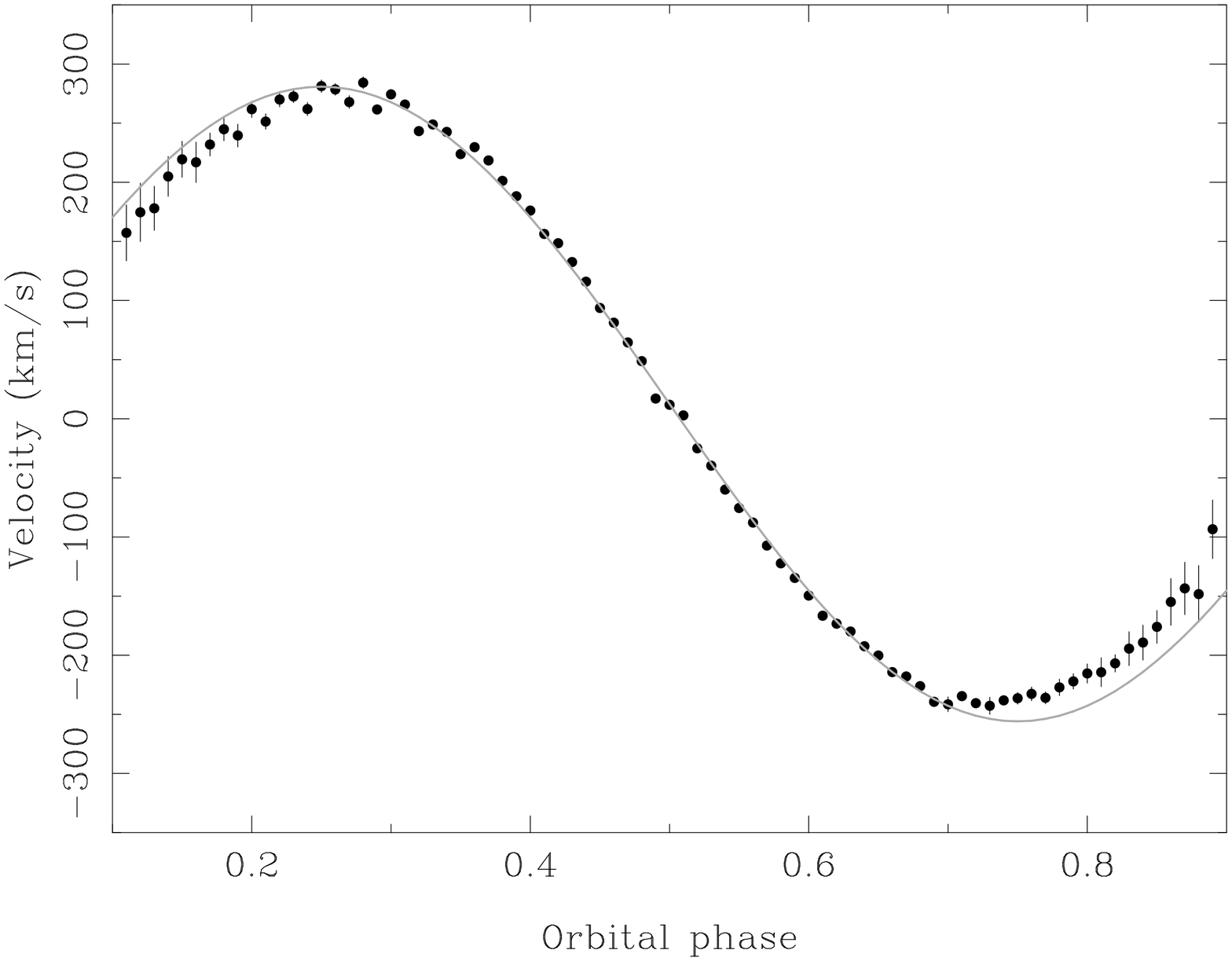}
    \includegraphics[width=0.3\textwidth]{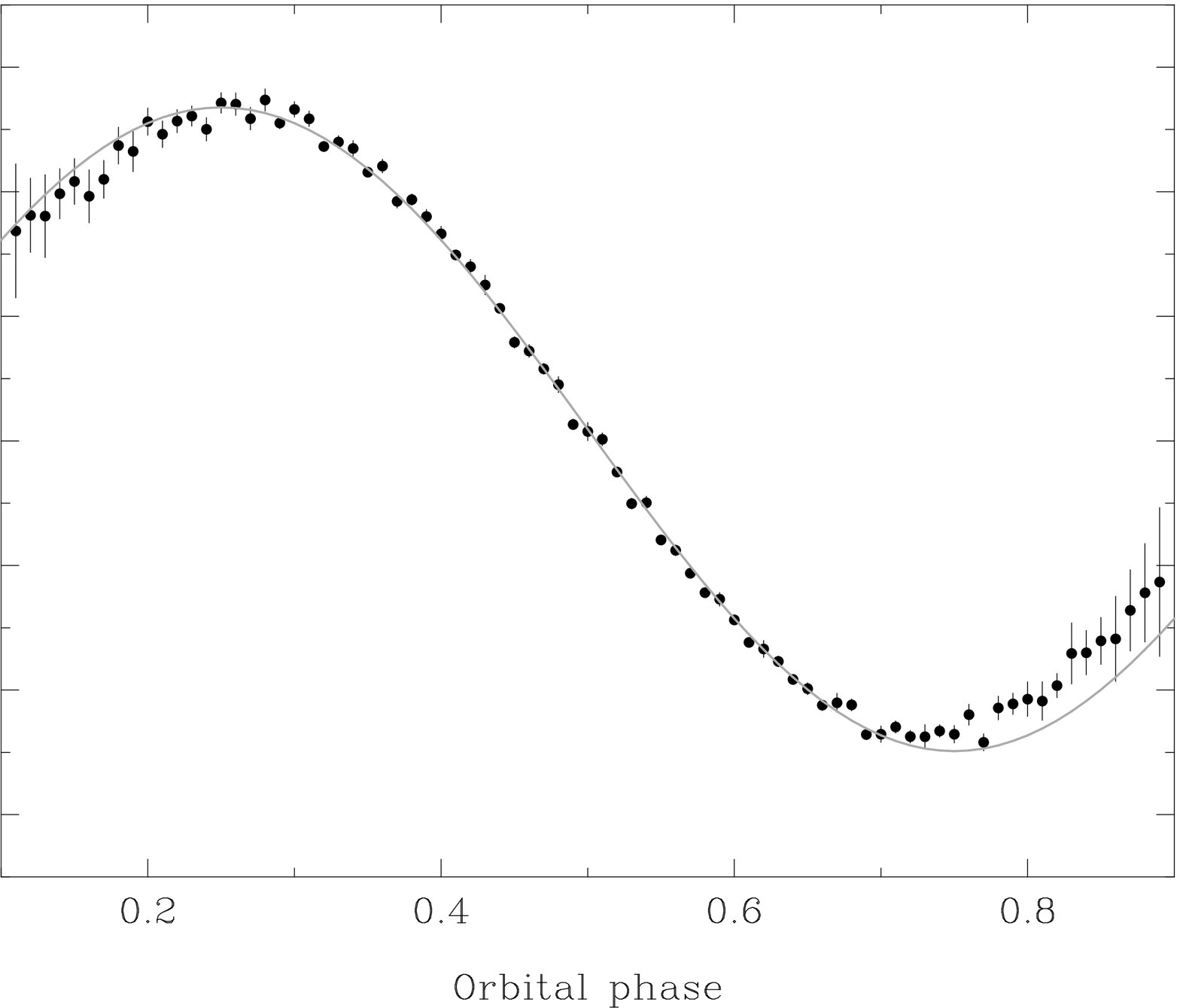}
    \includegraphics[width=0.3\textwidth]{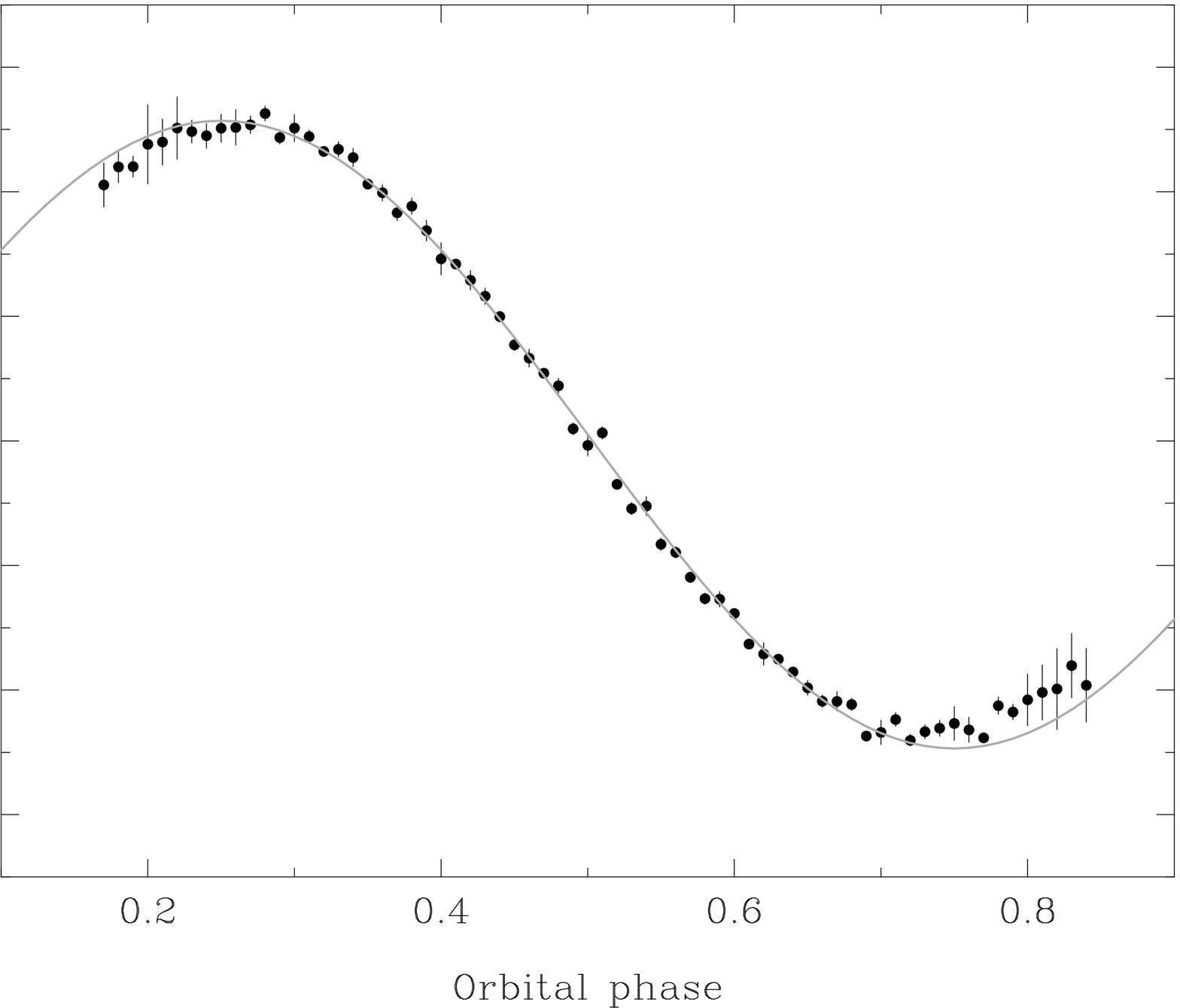} \\
  \caption{Sine curve fits for the Balmer lines (\emph{left}), the three strongest He I lines (\emph{centre}), and the MgII 4481\AA~ line (\emph{right}). The lines were fit using a straight line and a Gaussian. The MgII 4481\AA~ line becomes too faint before phase 0.15 and after phase 0.85 to fit.}
  \label{vfits}
  \end{center}
\end{figure*}

The spectra were shifted to the white dwarf's frame using the calculated radial velocity amplitude, then the spectra within the phase range quoted were averaged using weights for optimum signal-to-noise (ignoring those spectra taken during the eclipse). The result is the spectrum of the white dwarf component of NN Ser only, as shown in Figure~\ref{wdfull}. We match a homogeneously mixed hydrogen and helium atmosphere white dwarf model with a temperature of $57,000K$ and $\log{g}=7.5$ to the white dwarf spectrum (also shown in Figure~\ref{wdfull}). A helium abundance of $4.0\pm0.5\times10^{-4}$ by number is required to reproduce the measured equivalent width of the HeII line ($0.25\pm0.02$\AA). The result is in agreement with that of \citet{Haefner04} ($N_\mathrm{He} = 2 \pm 0.5 \times 10^{-4}$ by number) though there is some uncertainty in the treatment of Stark broadening in the code we used to calculate the model (TLUSTY,  \citealt{Hubeny88}, \citealt{Hubeny95}). The white dwarf spectrum shows only Balmer and HeII 4686\AA~ absorption features (the other sharp absorption features throughout the spectrum are interstellar absorption lines), no absorption lines are seen in the red spectra. This confirms previous results that the classification of the white dwarf is of type DAO1 according to the classification scheme of \citet{Sion83}.

\subsection{Secondary Star's Spectrum}

The most striking features of the UVES spectra are the emission lines arising from the heated face of the secondary star, the most prominent of which are the Balmer lines. Figure~\ref{trail} shows trailed spectra of several lines visible across the spectral range covered. The white dwarf component has been subtracted which creates a peak on the CaII trail that appears to move in anti-phase with the secondary, due to interstellar absorption. The top row shows three Balmer lines (H$\delta$,H$\gamma$ and H$\beta$) which clearly show reversed cores, becoming more prominent at increasing wavelength. Interestingly, the same effect is not visible in the Paschen series. Large broadening is obvious in the Hydrogen lines.

Radial velocities can be measured from these lines using the same multiple or single Gaussian plus polynomial approximations used to fit the white dwarf features. However, the measured radial velocity amplitude will be that of the emitting region on the face of the secondary star hence the radial velocity amplitude of the centre of mass of the secondary star will be larger than that measured from these lines (see Section 4.4). The white dwarf component shown in Figure~\ref{wdfull} was subtracted from each spectrum and the emission lines fitted. Due to the presence of interstellar absorption features, the subtraction of the white dwarf component creates peaks in the spectra since they show no phase variation. Figure~\ref{vfits} shows the fit to several lines: all the Balmer lines fitted simultaneously (H11 to H$\beta$), several HeI lines fitted simultaneously and a fit to the MgII 4481\AA~ line. All the lines show a similar deviation from a sinusoidal shape at small and large phases ($\la 0.25$ and $\ga 0.75$). This is because of the non-uniform distribution of flux over the secondary star. The radial velocity amplitude was calculated using only the points between these phases. The measured radial velocity amplitude varies for each line, the Balmer lines showing a larger radial velocity amplitude than most of the other lines. In addition, the radial velocity amplitude of the Balmer lines decreases towards the higher energy states. We believe that the spread in measured values is related to the optical depth for each line; this is discussed in Section 4.4. 

The Balmer, HeI and MgII lines from the UVES blue spectra were fitted with a polynomial and Gaussian. Furthermore, several HeI lines in the red spectra were fit as well as the Paschen lines (P12 to P$\epsilon$). Although several FeI lines are seen, they are too faint to calculate the radial velocity amplitude of the secondary star reliably. Nonetheless, they do show the same phase dependant variations as all the other emission lines. 

\begin{figure}
\begin{center}
 \includegraphics[width=\columnwidth]{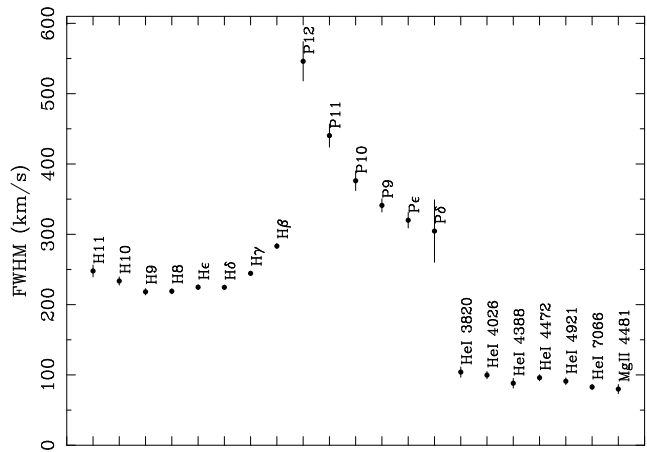}
 \caption{Widths of selected lines around the secondary eclipse, where they are at their widest. Note the large difference between the widths of the Hydrogen lines and that of every other species.}
 \label{line_widths}
\end{center}
\end{figure}

In addition to radial velocity information, the widths of the lines were fit. The widths of each of the emission lines vary strongly with phase and noticeable differences are seen between different atomic species. All of the hydrogen lines (both Balmer and Paschen lines) show a sharp increase in width around phase 0.1 which flattens off until phase 0.9 where the width falls off sharply. In contrast, the helium and magnesium lines show a gradual increase in width up to phase 0.5, then a gradual decrease. The behaviour of the hydrogen lines may be due to the lines saturating. Figure~\ref{line_widths} shows the average width of a selection of lines around the secondary eclipse (phase 0.5). The most striking feature is the width of the hydrogen lines which are at least double the width of any other line and reach widths of almost 600 km$\,$s$^{-1}$. An interesting trend is seen throughout the Balmer and Paschen series whereby the shorter wavelength lines are wider. The widths of these lines is probably an indication of Stark broadening which affects higher energy states to a larger extent. However, the longer wavelength Balmer lines become wider after H$\delta$ (the slight increase in the width of the H$\epsilon$ line is due to the overlapping CaII 3968\AA~ line and the nearby HeI 3965\AA~ line), presumably as a result of high optical depth leading to stronger saturation.

\begin{figure*}
\begin{center}
  \includegraphics[scale=0.35]{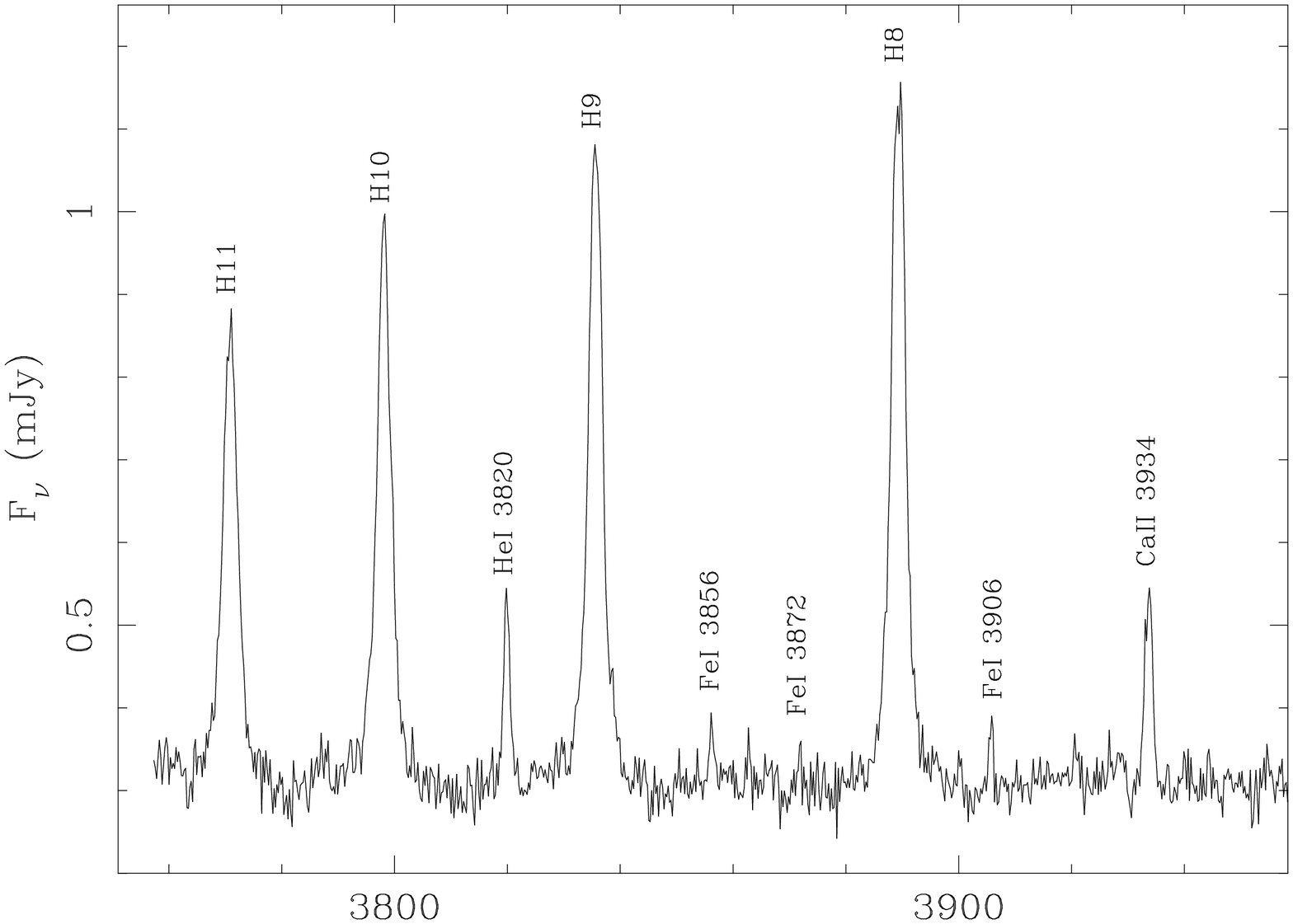} 
  \includegraphics[scale=0.35]{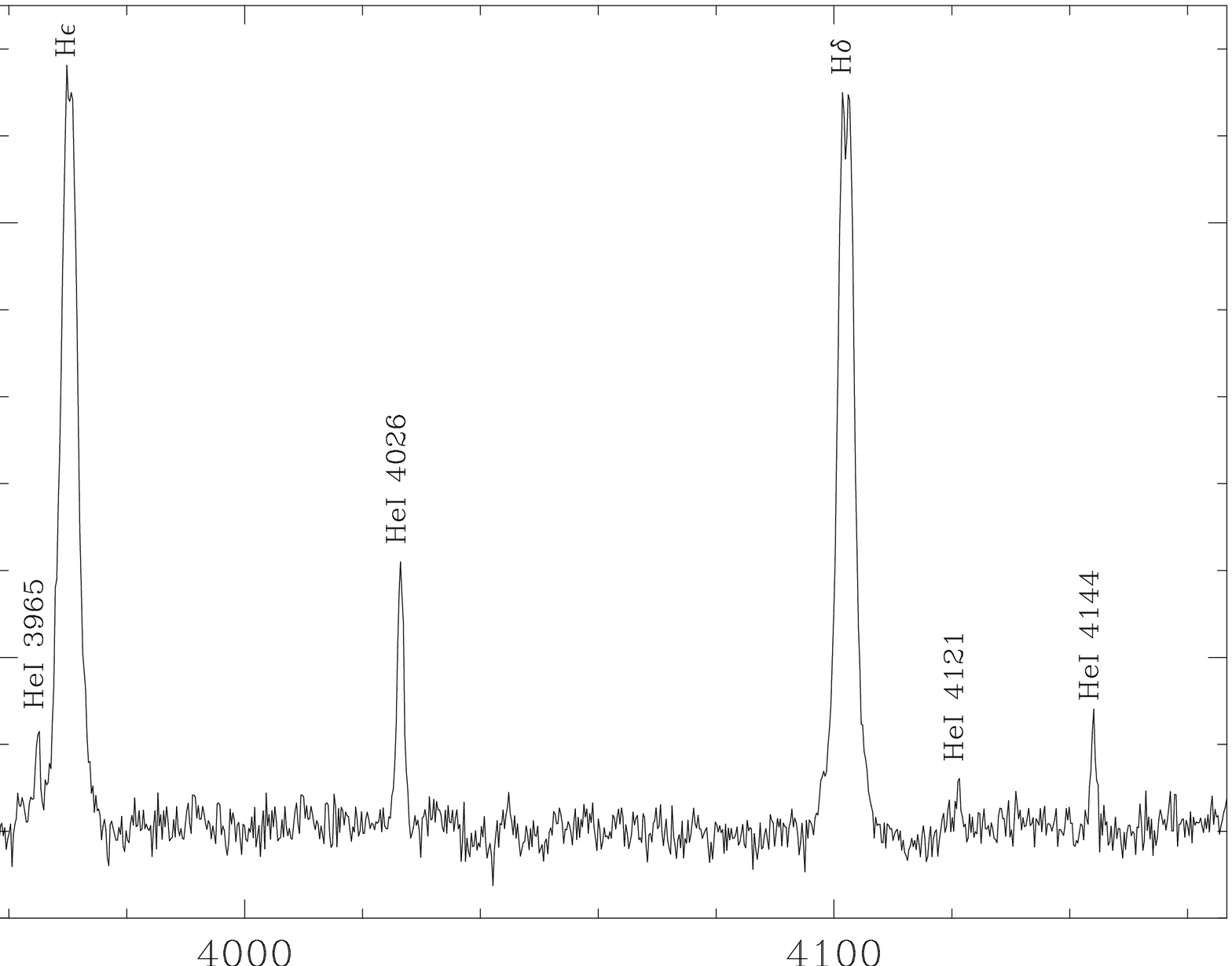} 
  \includegraphics[scale=0.35]{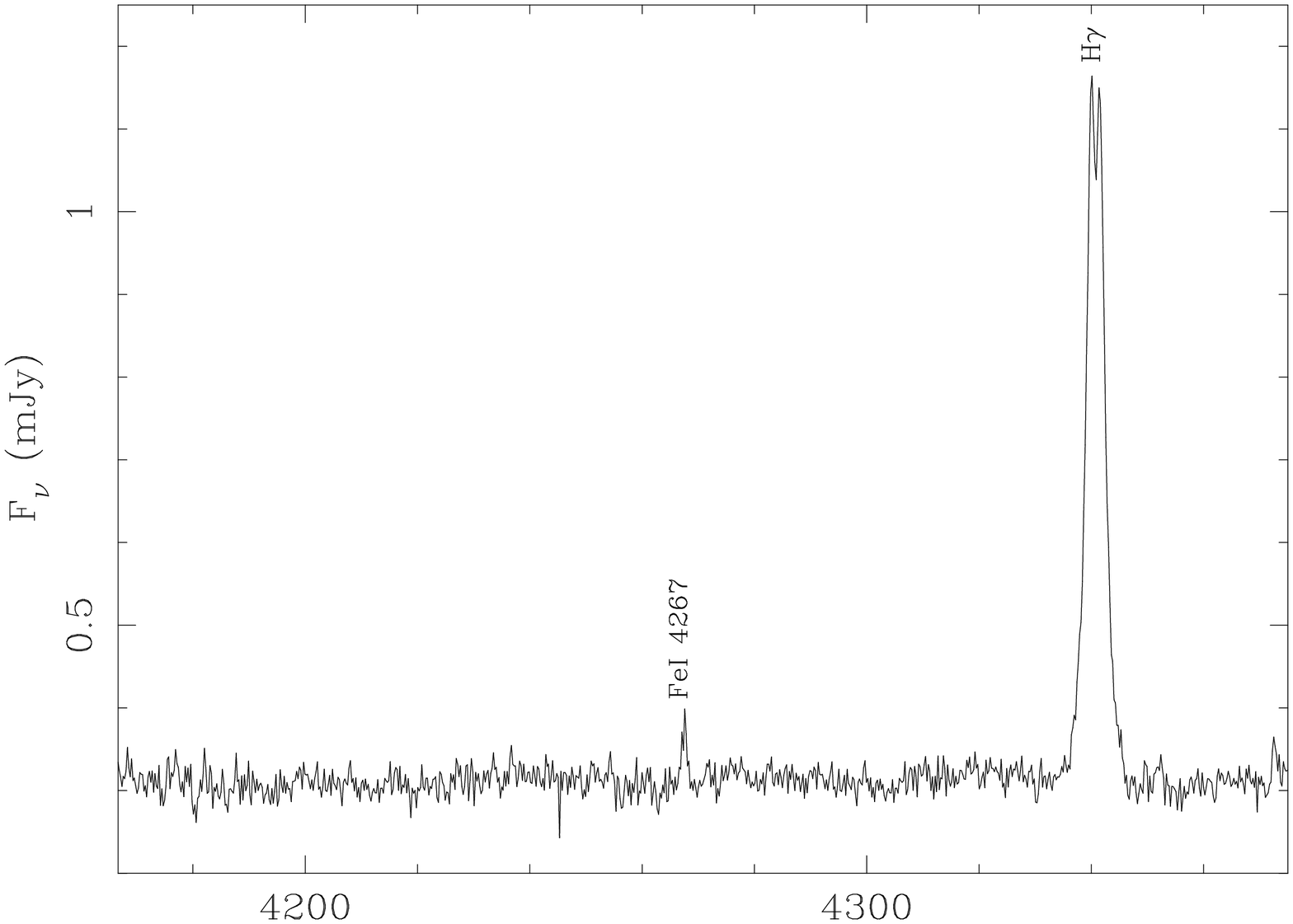} 
  \includegraphics[scale=0.35]{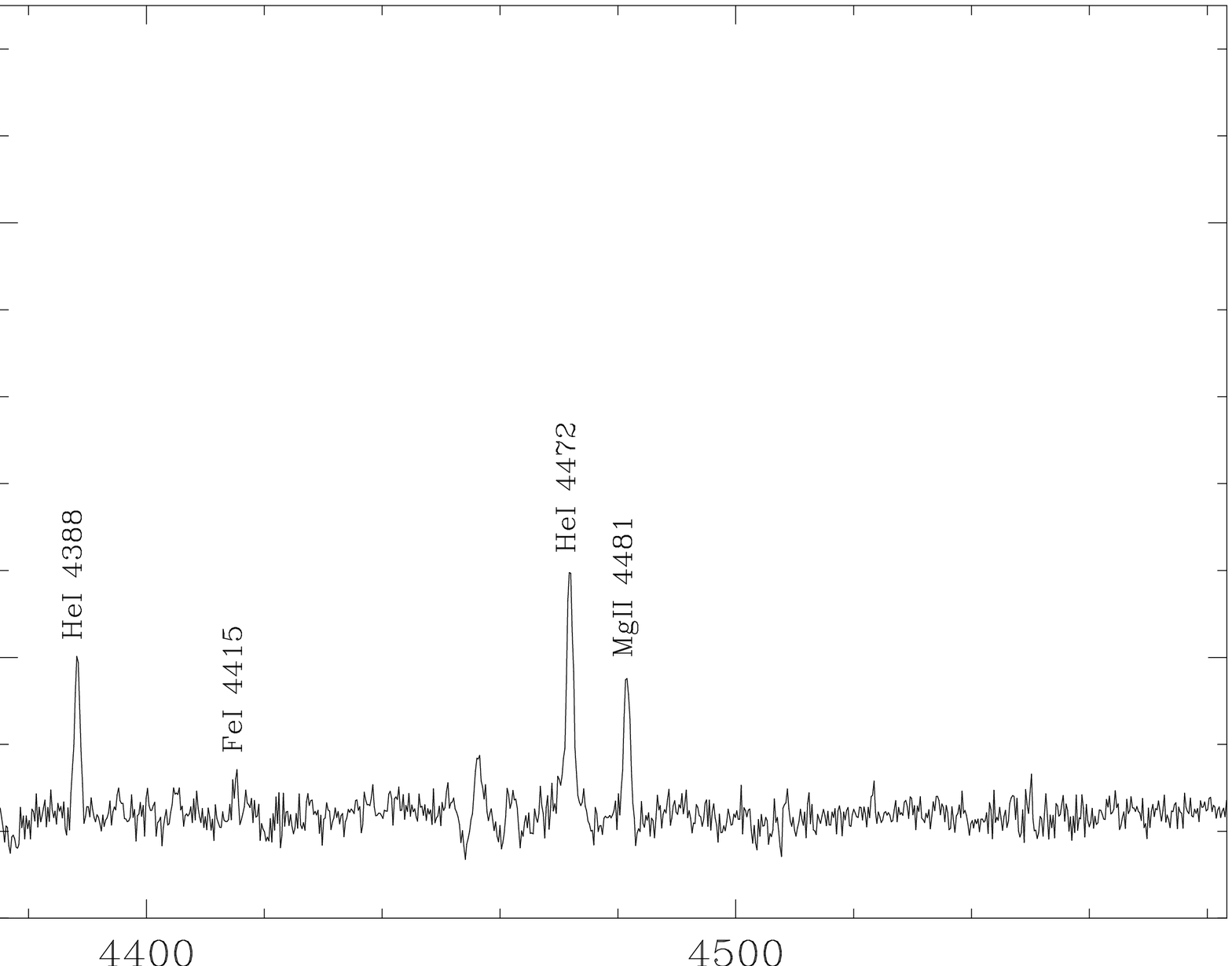}  
  \includegraphics[scale=0.35]{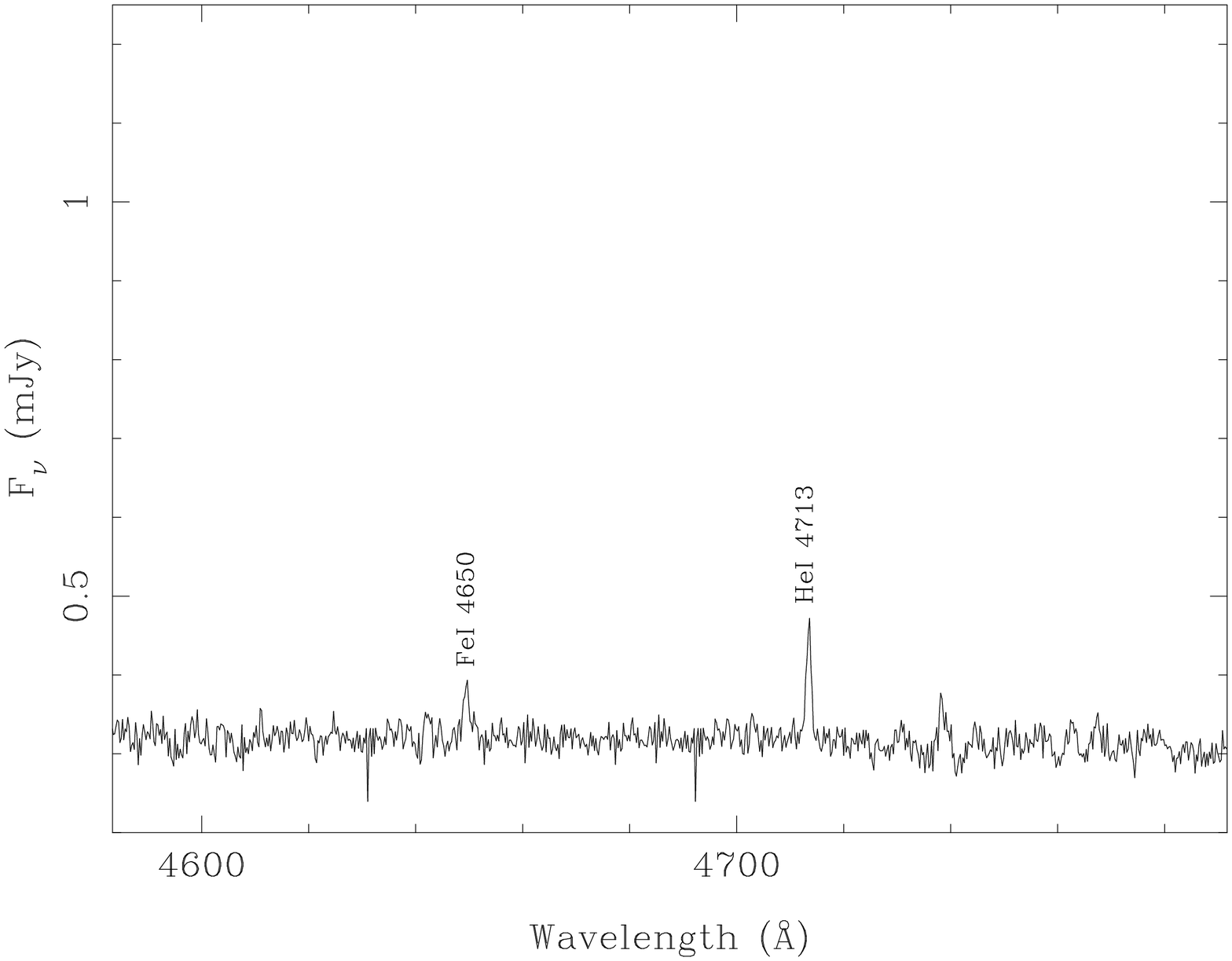} 
  \includegraphics[scale=0.35]{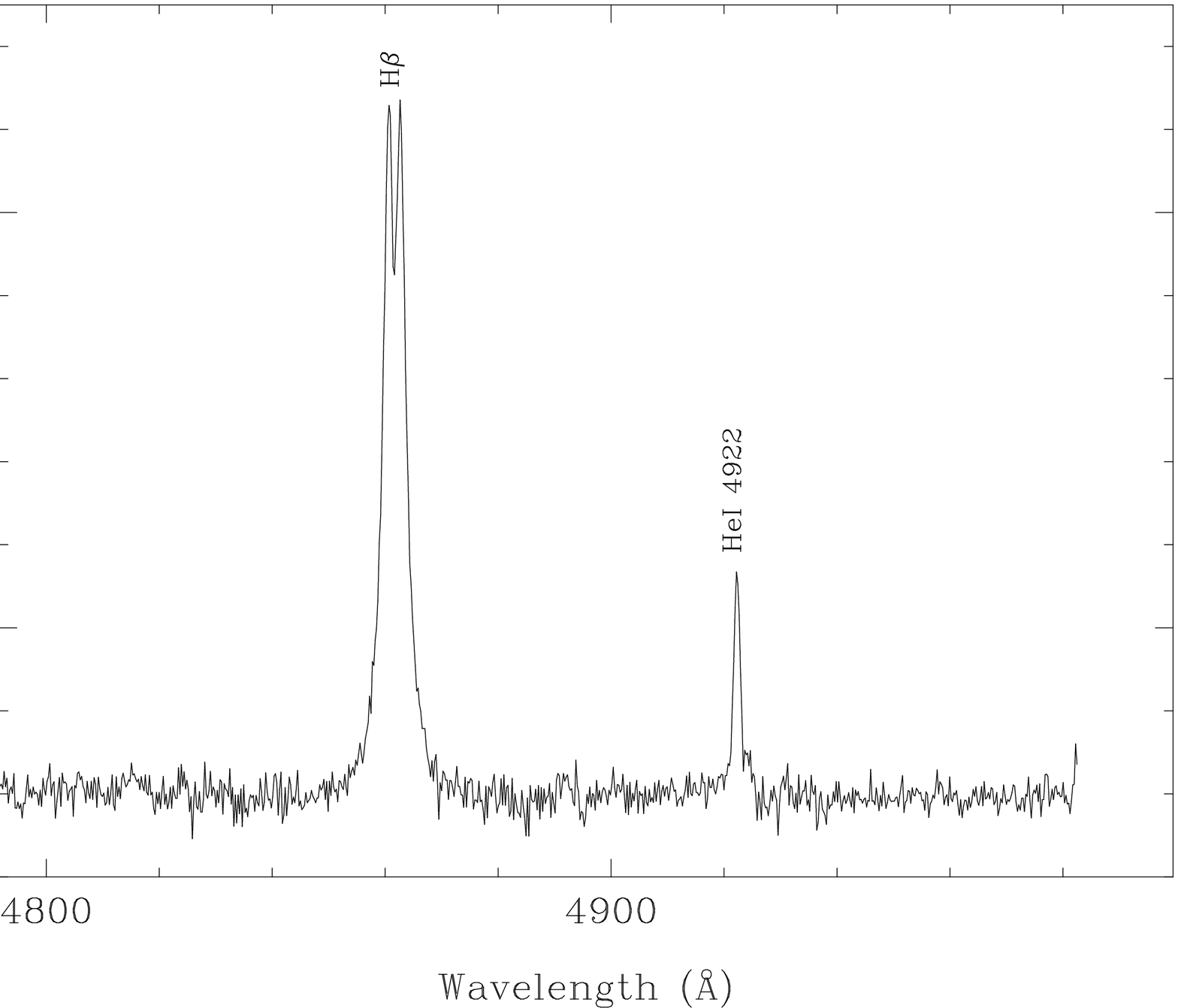} \\
 \caption{Blue spectrum of the heated part of the secondary star. The white dwarf component has been subtracted. The peaks seen just before the HeI 4472\AA~ line and after the HeI 4713\AA~ line are a result of interstellar absorption lines being subtracted off and are not real emission lines.}
 \label{secspec}
\end{center}
\end{figure*}

To recover the spectrum of the emitting region of the secondary star only, the simultaneous fit to all the HeI lines was used to shift the spectra to the frame of the emitting region of the secondary star. We use this radial velocity amplitude since it lies in the centre of the measured  amplitudes. The shifted spectra were then averaged, with larger weights given to those spectra taken around phase 0.5 where the reflection effect is greatest. Having previously subtracted the white dwarf's spectrum, the result is the spectrum of only the irradiated part of the secondary star. The UVES blue arm spectrum of the secondary star is shown in Figure~\ref{secspec} with the identified lines labelled. As previously mentioned, due to interstellar absorption lines, the subtraction of the white dwarf results in peaks that appear similar to the emission lines in Figure~\ref{secspec}. These occur just before the HeI 4472\AA~ line and after the HeI 4713\AA~ line. 

\begin{figure*}
\begin{center}
  \includegraphics[width=\columnwidth]{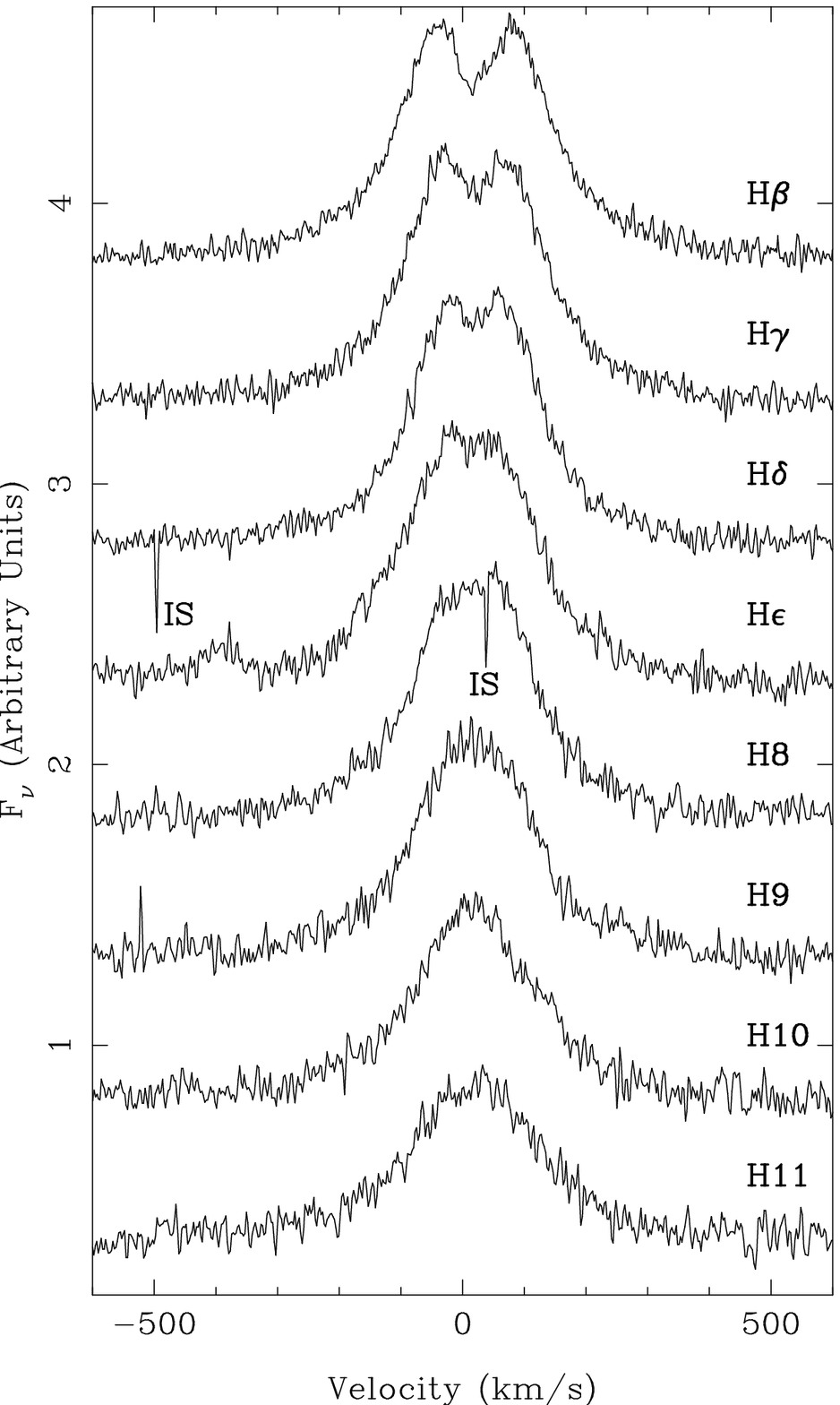}
  \includegraphics[width=\columnwidth]{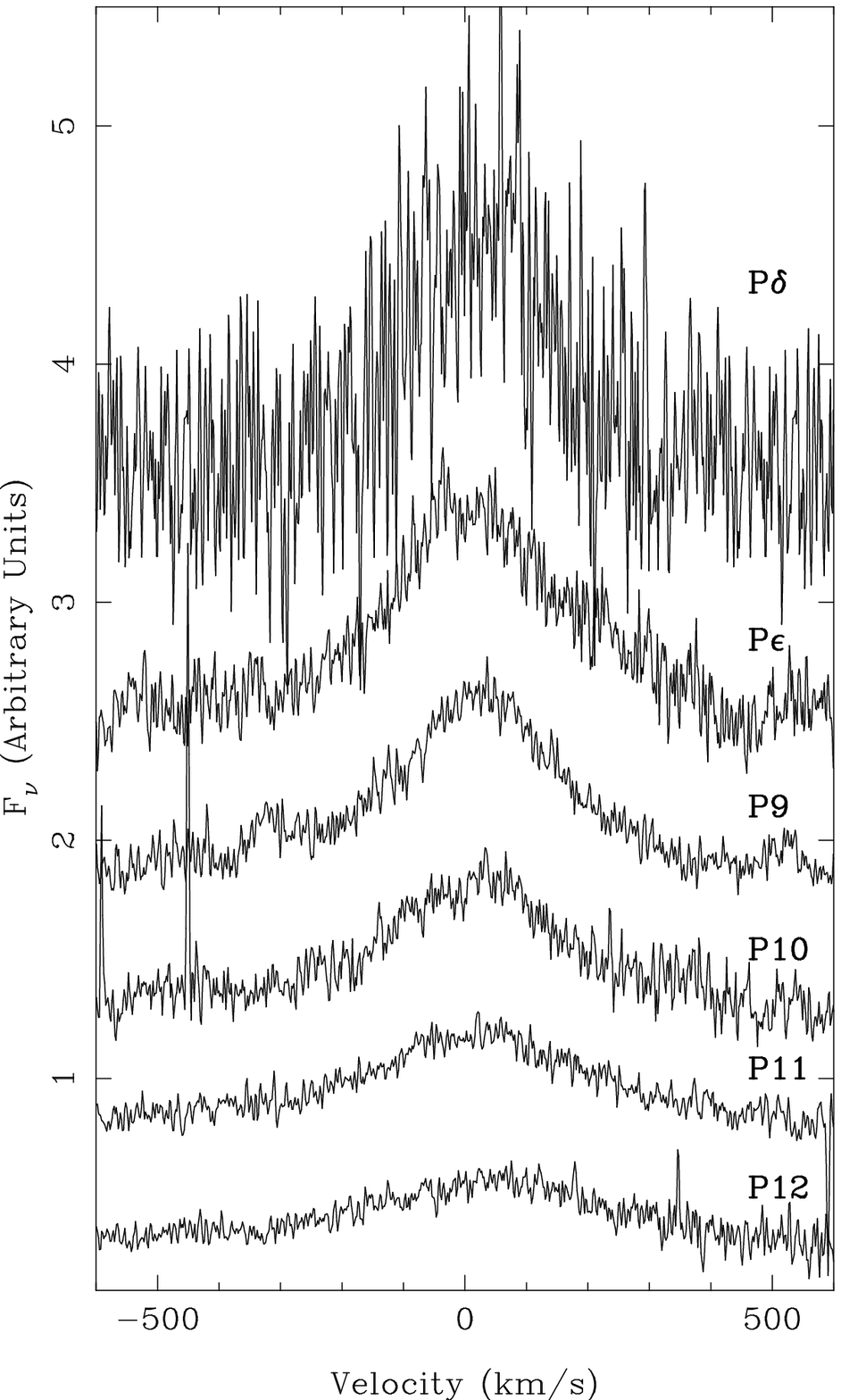}
 \caption{\emph{Left}: Profiles of the Balmer lines. IS corresponds to interstellar absorption features. \emph{Right}: Profiles of the Paschen lines. The white dwarf component has been subtracted and the motion of the secondary star removed.}
 \label{hydr_lines}
\end{center}
\end{figure*}

Once again the inverted core of the Balmer lines is the most striking feature. This can be seen even more clearly in Figure~\ref{hydr_lines}. For the shorter wavelength lines, there appears to be no core inversion. However, as the wavelength increases the effect becomes more pronounced. The separation between the two peaks in each line increases up to phase 0.5, reaching a maximum separation of 125 km$\,$s$^{-1}$ for the H$\beta$ line, the separation decreases symmetrically around phase 0.5. This effect was found to be caused by departure from local thermodynamic equilibrium (LTE) \citep{Barman04}. It is only seen in the hydrogen lines, and only the Balmer series. The Paschen lines, also shown in Figure~\ref{hydr_lines}, show no such core inversion. 

\begin{figure*}
\begin{center}
  \includegraphics[width=0.95\textwidth]{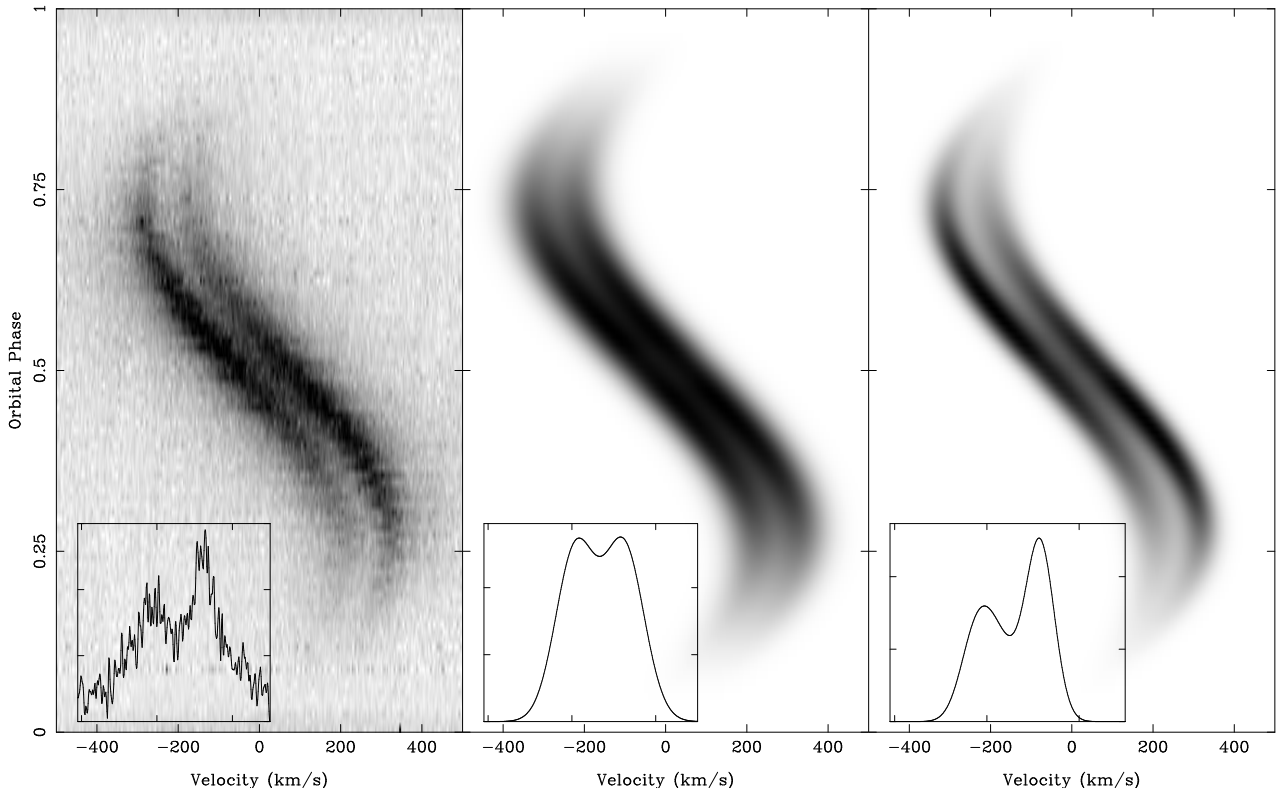}
 \caption{\emph{Left}: trailed spectrum of the H$\beta$ line with the white dwarf and continuum subtracted. \emph{Centre}: trailed spectrum of model line profiles convolved with two Gaussians set to reproduce the profile of the H$\beta$ emission line. \emph{Right}: trailed spectrum of a model with a varying line profile across the face of the secondary star, also set to reproduce the profile of the H$\beta$ emission line. Inset on each trail is the profile at phase 0.3; the first model fails to reproduce the observed asymmetry.}
 \label{invert}
\end{center}
\end{figure*}

The trailed spectra of the Balmer lines in Figure~\ref{trail} show, in addition to the core inversion, a clear asymmetry between the two peaks whereby the more shifted peak appears stronger (e.g. the most redshifted peak at phase 0.25 and the most blueshifted at phase 0.75). \citet{Barman04} found that the line profiles of the hydrogen Balmer lines in non-LTE have reversed cores caused by an over-population of the $n=2$ energy level. However, they did not take orbital effects into account. To test if the asymmetry is caused purely by non-LTE effects and orbital motion, we use a model line profile (see section 4.4 for details of the line profiles) and alter it such that the profile is double peaked across the surface of the secondary star. We do this by convolving the single peaked model with a pair of Gaussians over a full orbital phase, this mimics the reversed core behaviour whilst retaining the orbital effects. The width and separation of the Gaussians was set equal to that measured for the H$\beta$ line. A trailed spectrum of the resulting profile is shown in the central panel of Figure~\ref{invert} next to a trailed spectrum of the H$\beta$ line. The overall shape is similar and there is a small asymmetry, but the effect is far smaller than that seen in the H$\beta$ line, as seen in the profiles at phase 0.3 shown inset.

A possible explanation of the asymmetry is that the optical depth of the line emission varies across the surface of the secondary star. The emission originating from the heavily irradiated region is optically thick (producing the double peaked profile explained by \citealt{Barman04}). But further from this region, the irradiation flux decreases and the emission becomes more optically thin, changing the line profile towards a single peak. Since this region of optically thinner emission has a larger radial velocity amplitude, the more shifted peak's emission is increased resulting in the observed asymmetry.

To determine whether optical depth effects are responsible for the observed asymmetry, we adjust the model to vary the shape of the line profile across the face of the secondary star. We use a pair of Gaussians that get closer together as the irradiation flux decreases. The result is shown in the right hand panel of Figure~\ref{invert}. The model shows better agreement with the H$\beta$ line profile and the asymmetry is visible. Hence it is necessary to allow for the variation in irradiation levels and hence the non-LTE core reversal in order to model the line profiles in NN Ser.

Non-LTE effects have also been seen in other pre-cataclysmic variable systems such as HS 1857+5144 \citep{Aungwerojwit07} where the H$\beta$ and H$\gamma$ profiles are clearly double-peaked, V664 Cas and EC 11575--1845 \citep{Exter05} both show Stark broadened Balmer line profiles with absorption components. Likewise, double-peaked Balmer line profiles were observed in HS 1136+6646 \citep{Sing04} and Feige 24 \citep{Vennes94}; GD 448 \citep{Maxted98} also shows an asymmetry between the two peaks of the core-inverted Balmer lines. Since NN Ser is the only system observed with echelle resolution it shows this effect more clearly than any other system.

The secondary star's spectrum contains a large number of emission lines throughout. Each line was identified using rest wavelengths obtained from the National Institute of Standards and Technology\footnote[1]{http://physics.nist.gov/PhysRefData/ASD/lines\_form.html} (NIST) atomic spectra database. The velocity offset (difference between the rest and measured wavelength) and FWHM of each line were obtained by fitting with a straight line and a Gaussian. The line flux and equivalent width (EW) were also measured. Table~\ref{lines} contains a complete list of all the lines identified. In addition to the already known hydrogen, helium and calcium lines, there are a number of MgII lines throughout the spectra as well as FeI lines in the blue spectra and CI lines in the red spectra. The EW of the Balmer lines increases monotonically from H11 to H$\beta$ but the core inversion causes the line flux to level off after the H$\epsilon$ line. In addition to the Paschen P12 to P$\delta$ lines, half of the P13 line is seen (cut off by the spectral window used in the UVES upper red chip). 

The offsets for each line measured in Table~\ref{lines} combined with the measured offset of the HeII absorption line from the white dwarf, give a measurement of the gravitational redshift of the white dwarf. The HeII absorption line has an offset of $25.6\pm2.7$ km$\,$s$^{-1}$ measured in the same way as for the emission lines. Taking a weighted average of the emission line velocities gives an offset of $15.1\pm0.2$ km$\,$s$^{-1}$, resulting in a gravitational redshift of $10.5\pm2.7$ km$\,$s$^{-1}$ for the white dwarf. This fairly small redshift is most likely due to the fact that the white dwarf is very hot leading to an inflated radius, as we will see shortly.

\section{System Parameters}
\subsection{Light Curve Analysis} 

Analysis of the ULTRACAM light curves gives strong constraints on the system parameters. A code written to produce models for the general case of binaries containing a white dwarf was used (see \citealt{Copperwheat09}, submitted, for details). It has been used in the study of other white dwarf-main sequence binaries \citep{Pyrzas09}. Several components of the model include accretion phenomena for the analysis of cataclysmic variables. Since NN Ser is a detached system these components were not included. The program subdivides each star into small elements with a geometry fixed by its radius as measured along the direction of centres towards the other star. Roche geometry distortion and irradiation of the secondary star are included, the irradiation is approximated by $\sigma T^{\prime}{}_\mathrm{sec}^{4}= \sigma T_\mathrm{sec}^{4}+ F_\mathrm{irr}$ where $T^{\prime}{}_\mathrm{sec}$ is the modified temperature and $T_\mathrm{sec}$ the temperature of the unirradiated companion, $\sigma$ is the Stefan-Boltzmann constant and $F_\mathrm{irr}$ is the irradiating flux, accounting for the angle of incidence and distance from the white dwarf. 

From an initial set of parameters defined by the user, the code produces model light curves which are initially fitted to the ULTRACAM data using Levenberg-Marquardt minimisation \citep{Press86} to produce a set of covariances. The resultant model parameters are fitted to the ULTRACAM data using Markov chain Monte Carlo (MCMC) minimisation, using the covariances from the Levenberg-Marquardt minimisation to define the parameter jumps, to produce the final parameters and their errors; we follow the procedures described in \citet{Collier07}.

The parameters needed to define the model are: the mass ratio, $q = M_\mathrm{sec}/M_\mathrm{WD}$, the inclination, $i$, the radii scaled by the binary separation, $R_\mathrm{WD}/a$ and $R_\mathrm{sec}/a$, the unirradiated temperatures, $T_\mathrm{eff,WD}$ and $T_\mathrm{eff,sec}$, linear limb darkening coefficients for the white dwarf and secondary star, the time of mid eclipse, $T_{0}$, the period, $P$, the gravity darkening coefficient for the secondary star and the fraction of the irradiating flux from the white dwarf absorbed by the secondary star.

\begin{table*}
 \centering
 \begin{minipage}{140mm}
  \centering
  \caption{Identified emission lines in the UVES spectra. Each line was fitted with a Gaussian to determine the velocity and FWHM.}
  \label{lines}
  \begin{tabular}{@{}lccccl@{}}
  \hline
   Line ID&Velocity&FWHM&Line Flux ($10^{-15}$&Equivalent& comment\\
  &(km$\,$s$^{-1}$)&(km$\,$s$^{-1}$)&ergs$\,$cm$^{-2}$s$^{-1}$\AA$^{-1}$)&Width (\AA)& \\
 \hline
  H11 3770.634        & 16.4$\pm$1.5&250.7$\pm$4.1&3.10(4)&4.36(5) & - \\ 
  H10 3797.910        & 12.6$\pm$1.1&233.9$\pm$2.9&3.46(4)&4.74(5) & - \\
  HeI 3819.761        & 12.4$\pm$1.8&107.4$\pm$4.5&0.46(2)&0.69(3) & - \\
  H9 3835.397         & 13.5$\pm$0.9&223.1$\pm$2.4&4.32(4)&6.30(6) & - \\
  FeI 3856.327        & 11.0$\pm$3.7& 63.2$\pm$9.9&0.12(1)&0.20(2) & - \\
  FeI 3871.749        & 14.7$\pm$2.8& 50.0$\pm$7.0&0.11(2)&0.20(3) & - \\
  H8 3889.055         & 14.6$\pm$0.7&223.1$\pm$1.8&4.46(3)&6.51(4) & - \\
  FeI 3906.479        & 10.1$\pm$3.6& 68.2$\pm$8.8&0.22(2)&0.41(4) & - \\
  CaII 3933.663       & -2.9$\pm$1.4&106.8$\pm$3.7&0.64(2)&1.06(3) & Interstellar absorption present \\
  HeI 3964.727        & 12.2$\pm$3.2& 85.7$\pm$8.0&0.30(2)&0.55(3) & Close to the H$\epsilon$ line \\
  H$\epsilon$ 3970.074&  4.6$\pm$0.7&249.4$\pm$2.0&5.44(3)&9.59(6) & - \\
  HeI 4026.189        & 14.3$\pm$1.0& 96.3$\pm$2.8&0.71(2)&1.28(3) & - \\
  H$\delta$ 4101.735  & 17.1$\pm$0.6&231.0$\pm$1.5&4.72(3)&8.57(5) & - \\
  HeI 4120.824        & 10.3$\pm$4.4& 73.9$\pm$9.8&0.12(1)&0.24(2) & - \\
  HeI 4143.759        & 15.6$\pm$2.5& 85.2$\pm$6.4&0.29(1)&0.59(3) & - \\
  FeI 4266.964        & 12.8$\pm$4.5& 97.9$\pm$10.5&0.22(2)&0.49(3) & - \\
  H$\gamma$ 4340.465  & 14.5$\pm$0.6&255.9$\pm$1.5&4.74(2)&9.40(4) & - \\
  HeI 4387.928        & 18.1$\pm$1.4& 80.7$\pm$3.6&0.34(1)&0.74(2) & - \\
  FeI 4415.122        & 15.5$\pm$4.4& 66.4$\pm$11.1&0.18(1)&0.42(2) & - \\
  HeI 4471.681        & 14.5$\pm$1.0& 86.3$\pm$2.5&0.54(2)&1.15(2) & - \\
  MgII 4481.327       & 16.9$\pm$1.3& 89.5$\pm$3.5&0.33(1)&0.74(2) & - \\
  FeI 4649.820        & 18.9$\pm$4.6& 80.7$\pm$8.8&0.13(1)&0.30(2) & - \\
  HeI 4713.146        & 14.7$\pm$1.4& 76.6$\pm$3.7&0.12(1)&0.57(2) & - \\
  H$\beta$ 4861.327   & 14.7$\pm$0.4&262.0$\pm$1.3&4.26(2)&10.34(4) & - \\
  HeI 4921.929        & 16.4$\pm$1.0& 81.7$\pm$2.5&0.39(1)&1.05(2) & - \\
  HeI 7065.709        & 15.4$\pm$0.4& 78.4$\pm$1.1&0.27(3)&1.17(1) & - \\
  HeI 7281.349        & 16.0$\pm$0.7& 67.3$\pm$2.0&0.13(3)&0.64(1) & - \\
  MgII 7877.051       & 12.6$\pm$3.1& 37.1$\pm$7.8&0.12(3)&0.23(2) & - \\
  MgII 7896.368       & 12.9$\pm$2.2& 54.8$\pm$5.7&0.15(4)&0.30(2) & - \\
  CI 8335.15          &  9.9$\pm$4.1& 44.2$\pm$6.6&0.11(4)&0.20(3) & - \\
  CaII 8498.02        & 14.4$\pm$2.7& 99.8$\pm$8.3&0.14(5)&0.90(3) & - \\
  P12 8750.473        & 16.6$\pm$5.3&402.4$\pm$16.8&1.15(2)&5.49(1) & Half of P13 line seen as well \\
  P11 8862.784        & 16.5$\pm$2.7&376.2$\pm$8.0&1.68(2)&8.16(9) & - \\
  CaII 8927.36        & 18.0$\pm$3.6& 58.3$\pm$9.5&0.13(1)&0.65(5) & - \\
  P10 9014.911        & 13.1$\pm$2.7&280.4$\pm$8.1&1.97(2)&10.27(9) & - \\
  CI 9061.43          & 12.1$\pm$4.6& 35.6$\pm$7.7&0.12(3)&0.28(4) & - \\
  MgII 9218.248       & 16.2$\pm$2.6& 48.9$\pm$7.6&0.12(6)&0.51(4) & Close to the P9 line\\
  P9 9229.015         & 15.1$\pm$1.3&316.5$\pm$3.6&2.42(2)&11.47(8)& - \\
  MgII 9244.266       & 16.0$\pm$2.5& 44.6$\pm$6.4&0.14(3)&0.40(3) & -\\
  CI 9405.73          & 10.7$\pm$5.1& 46.0$\pm$8.0&0.15(1)&0.87(7) & -\\
  P$\epsilon$ 9545.972& 16.4$\pm$2.3&332.3$\pm$6.4&3.39(4)&19.58(7) & - \\
  P$\delta$ 10049.374 & 18.4$\pm$7.4&253.4$\pm$22.2&2.7(1)&18.0(7) & Very noisy in the far red \\
\hline
\end{tabular}
\end{minipage}
\end{table*}

After a preliminary estimate we kept the mass ratio fixed at 0.2 (the light curves are only weakly dependent on this parameter), the temperature of the white dwarf fixed at 57,000K \citep{Haefner04}, the gravity darkening coefficient fixed at 0.08 (the usual value for a convective atmosphere) and the limb darkening coefficient of the white dwarf fixed at different values for each filter based on a white dwarf with $T_\mathrm{eff} = 57,000$K and $\log{g} = 7.46$ using ULTRACAM \emph{u'g'r'i'z'} filters \citep{Gansicke95}. The $\log{g}$ was obtained from an initial MCMC minimisation of the \emph{g'} light curve with the limb darkening coefficient of the white dwarf fixed at a value of 0.2 (the $\log{g}$ determined from this is very similar to the final value determined later). All other parameters were optimised in the MCMC minimisation. The initial values for the inclination, radii and the temperature of the secondary star were taken from \citet{Haefner04}, the limb darkening coefficient for the secondary star was initially set to zero and the fraction of the irradiating flux from the white dwarf absorbed by the secondary star was initially set to 0.5 (note that the intrinsic flux of the secondary star is negligible).

\begin{table*}
 \centering
  \caption{Best fit parameters from Markov chain Monte Carlo minimisation for each ULTRACAM light curve. Lin\_limb is the linear limb darkening coefficient for the white dwarf which was kept fixed, the values quoted are for a model white dwarf of temperature 57,000K and $\log{g} = 7.46$. Absorb is the fraction of the irradiating flux from the white dwarf absorbed by the secondary star.}
  \label{mcmcfit}
  \begin{tabular}{@{}lcccc@{}}
  \hline
 Parameter & \emph{u'}& \emph{g'}& \emph{r'}& \emph{i'} \\
 \hline 
 Inclination             & $89.18 \pm 0.27$      & $89.67 \pm 0.05$      & $89.31 \pm 0.21$      & $89.59 \pm 0.27$      \\
 $R_\mathrm{WD}/a$        & $0.02262 \pm 0.00014$ & $0.02264 \pm 0.00002$ & $0.02271 \pm 0.00010$ & $0.02257 \pm 0.00010$ \\
 $R_\mathrm{sec}/a$       & $0.1660 \pm 0.0011$   & $0.1652 \pm 0.0001$   & $0.1657 \pm 0.0007$   & $0.1654 \pm 0.0003$   \\
 $T_\mathrm{sec}$         & $3962 \pm 32$         & $3125 \pm 10$         & $3108 \pm 11$         & $3269 \pm 7$          \\
 Lin\_limb$_\mathrm{WD}$  & 0.125                 & 0.096                 & 0.074                 & 0.060                 \\
 Lin\_limb$_\mathrm{sec}$ & $-1.44 \pm 0.13$      & $-0.48 \pm 0.03$      & $-0.26 \pm 0.02$      & $-0.06 \pm 0.03$      \\
 Absorb                  & $0.899 \pm 0.001$     & $0.472 \pm 0.001$     & $0.604 \pm 0.006$     & $0.651 \pm 0.005$     \\
 \hline
\end{tabular}
\end{table*} 

\begin{figure*}
\begin{center}
 \includegraphics[width=0.999\columnwidth]{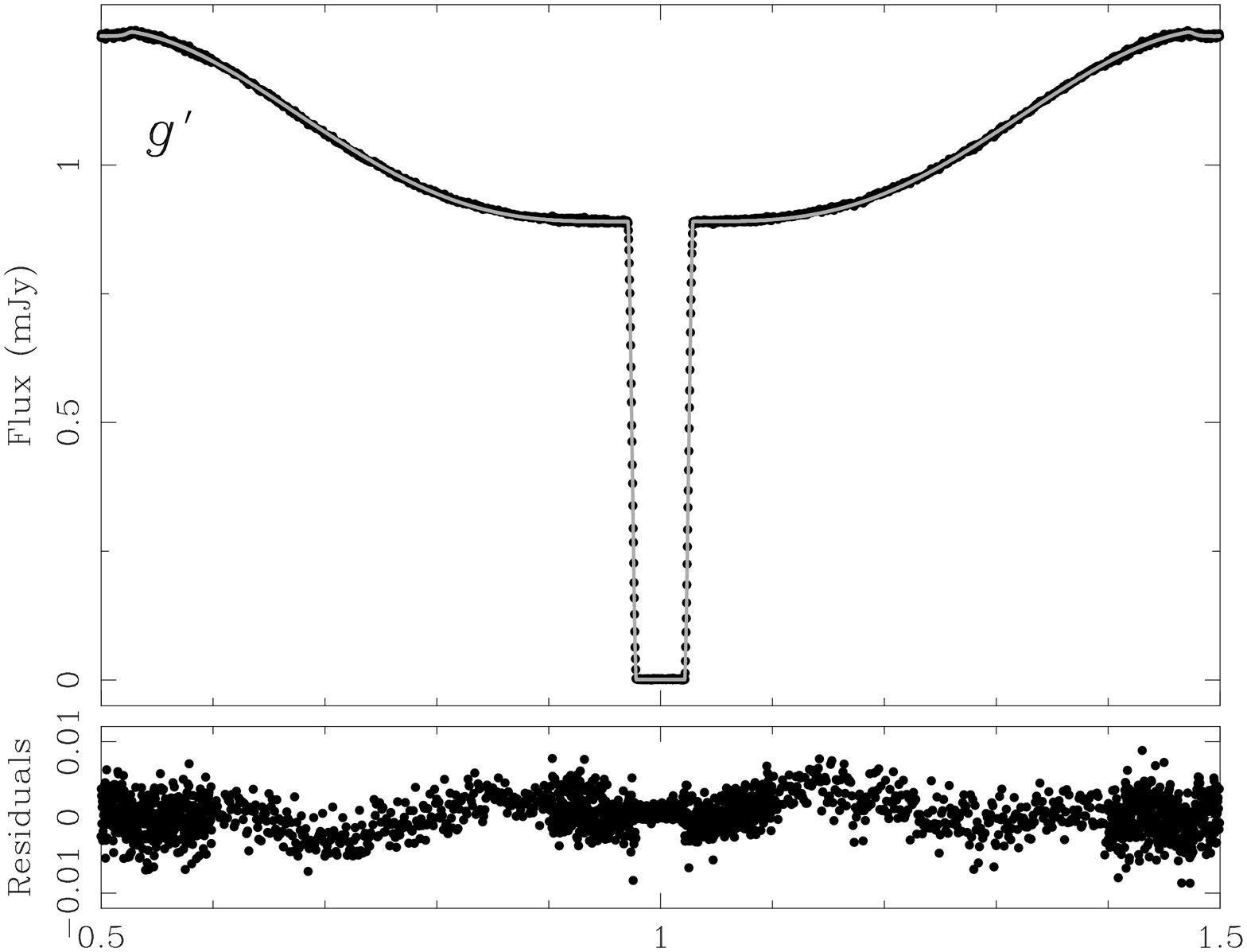}
 \includegraphics[width=0.963\columnwidth]{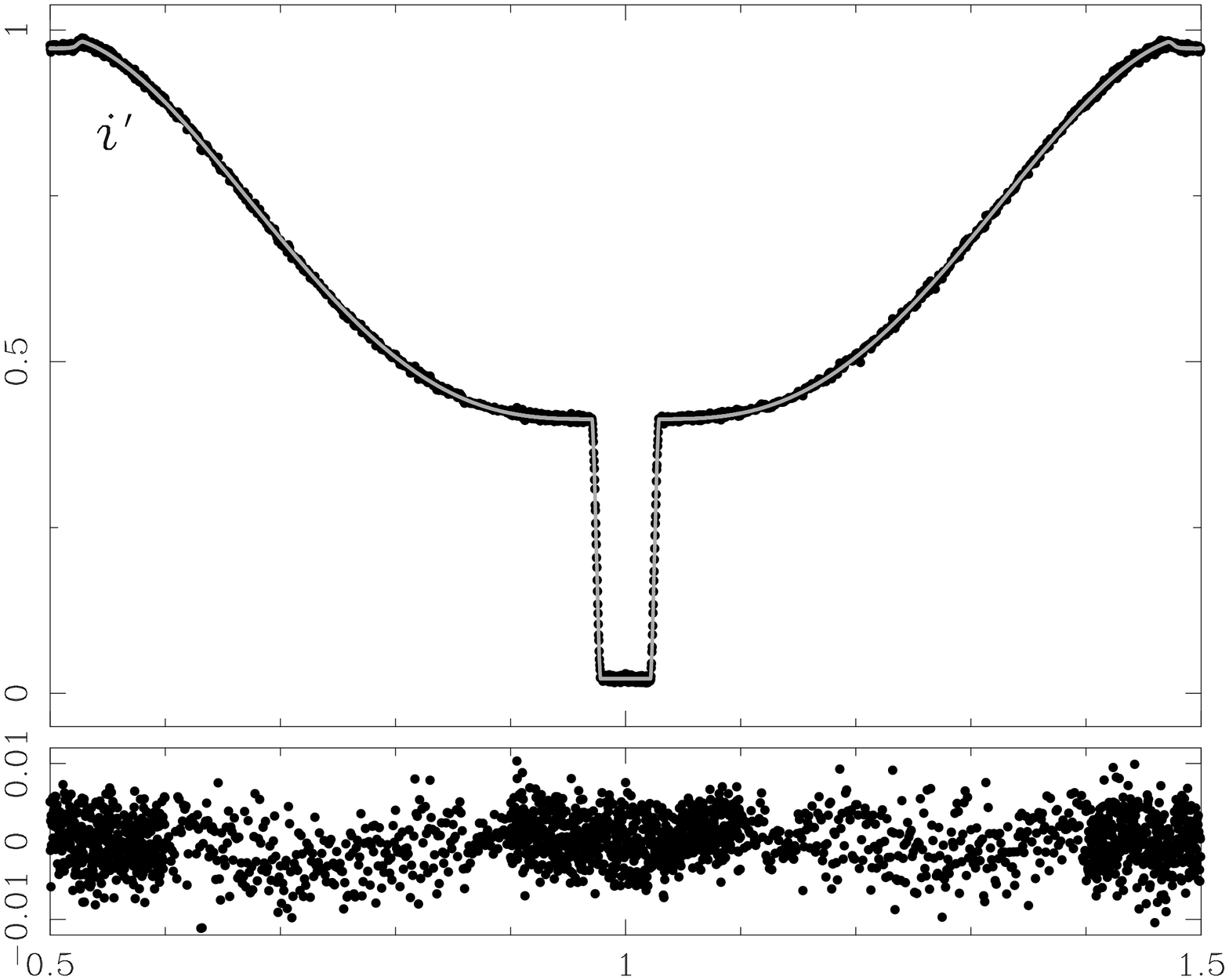} \\
 \includegraphics[width=0.315\textwidth]{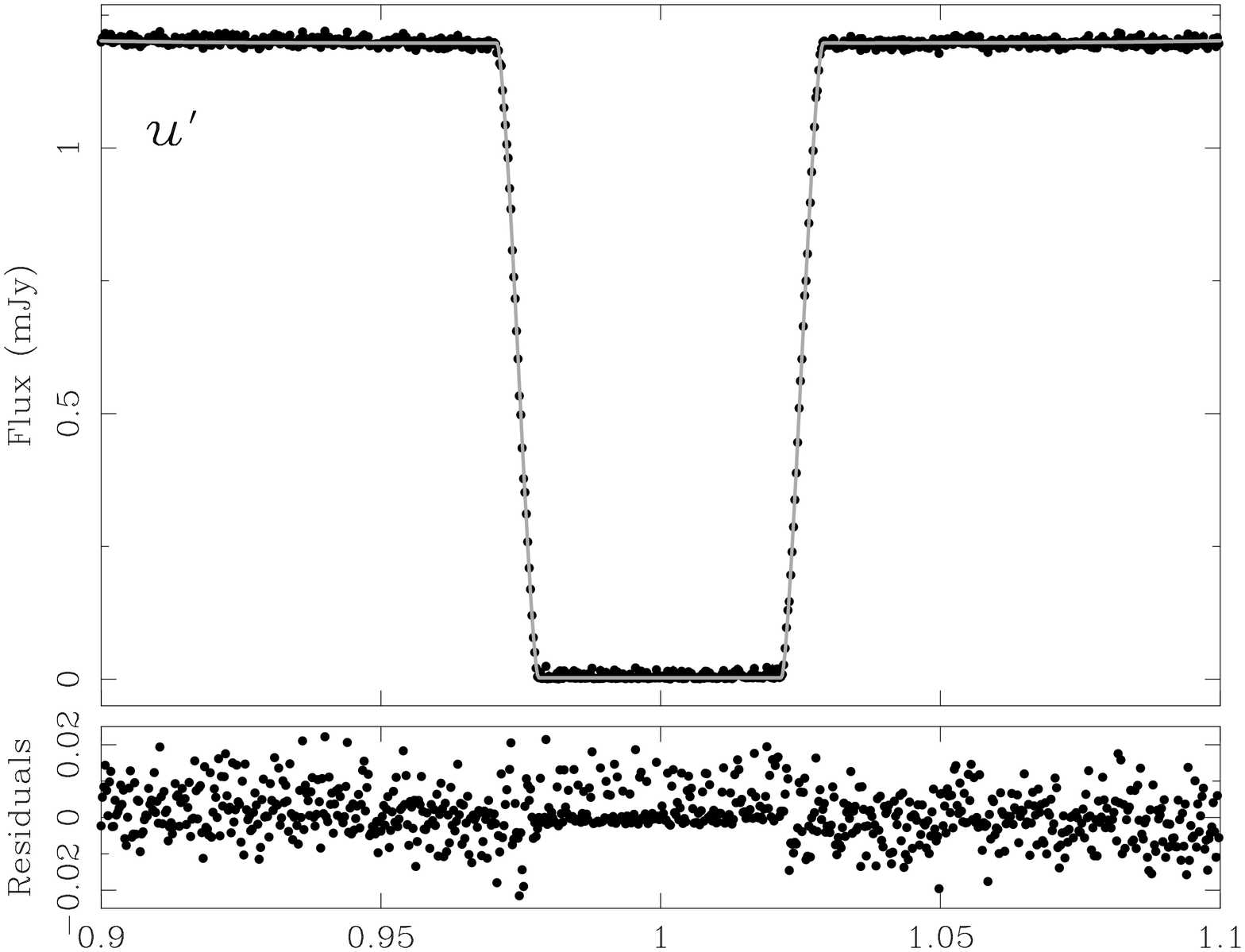}
 \includegraphics[width=0.3\textwidth]{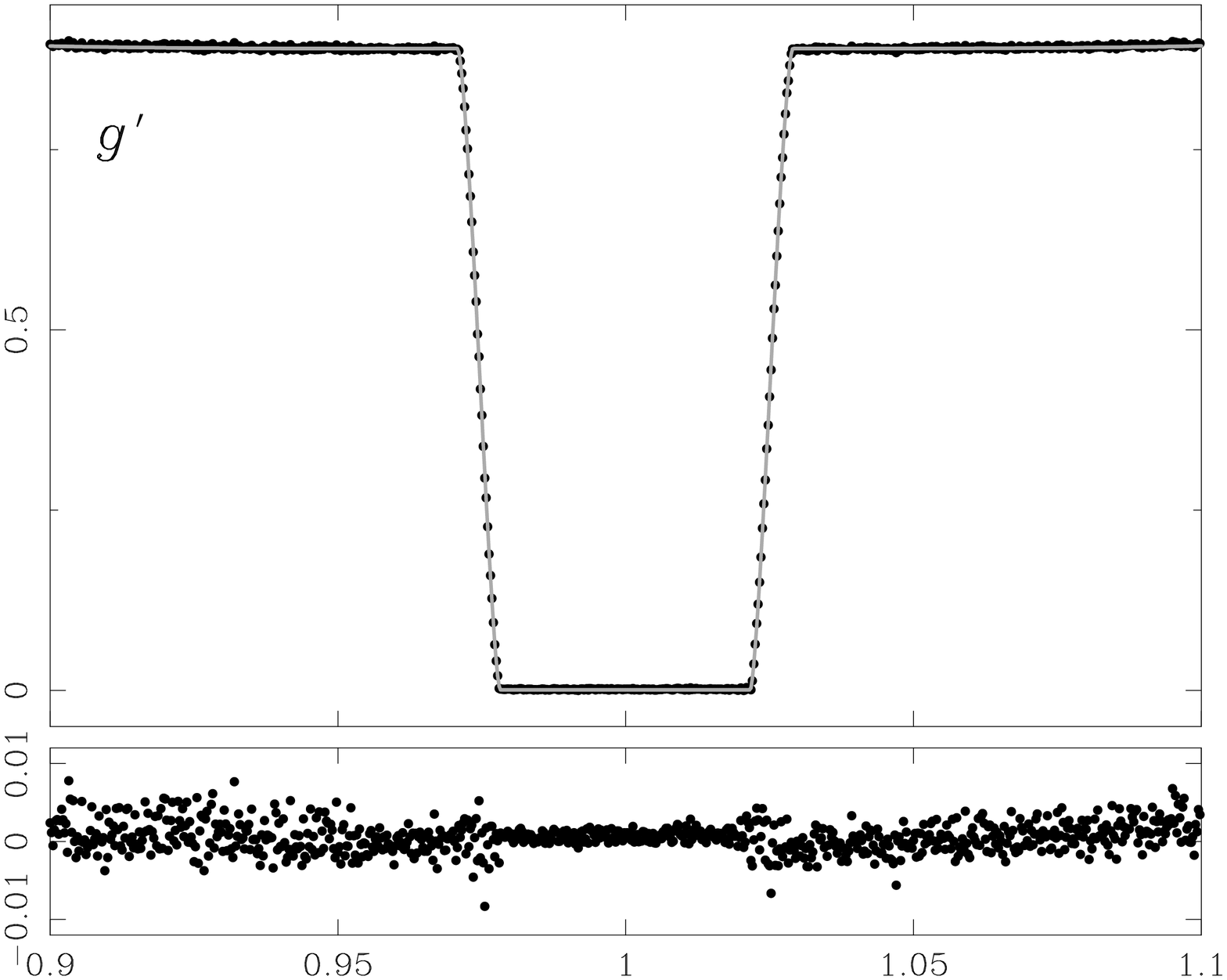}
 \includegraphics[width=0.3\textwidth]{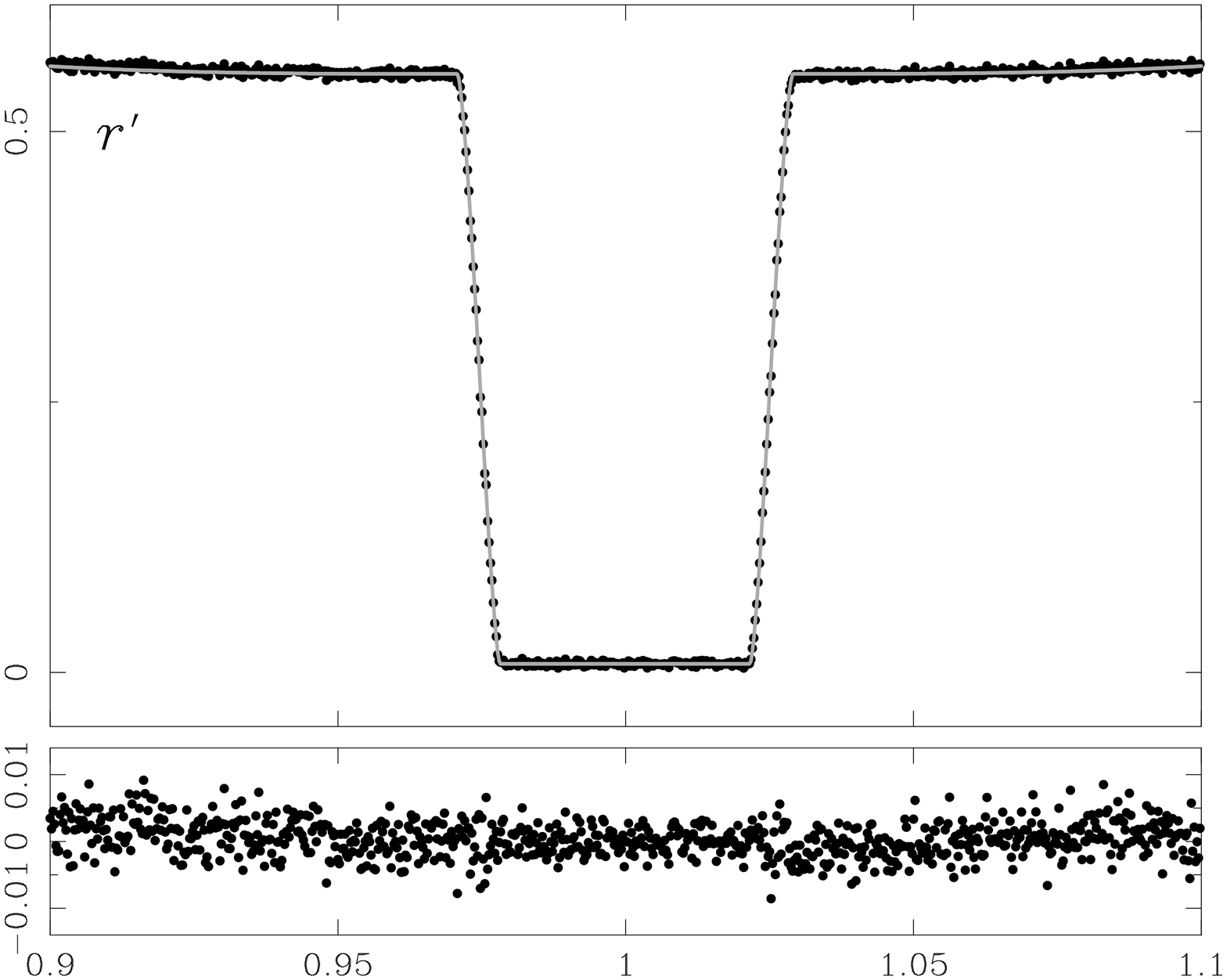} \\
 \includegraphics[width=0.31\textwidth]{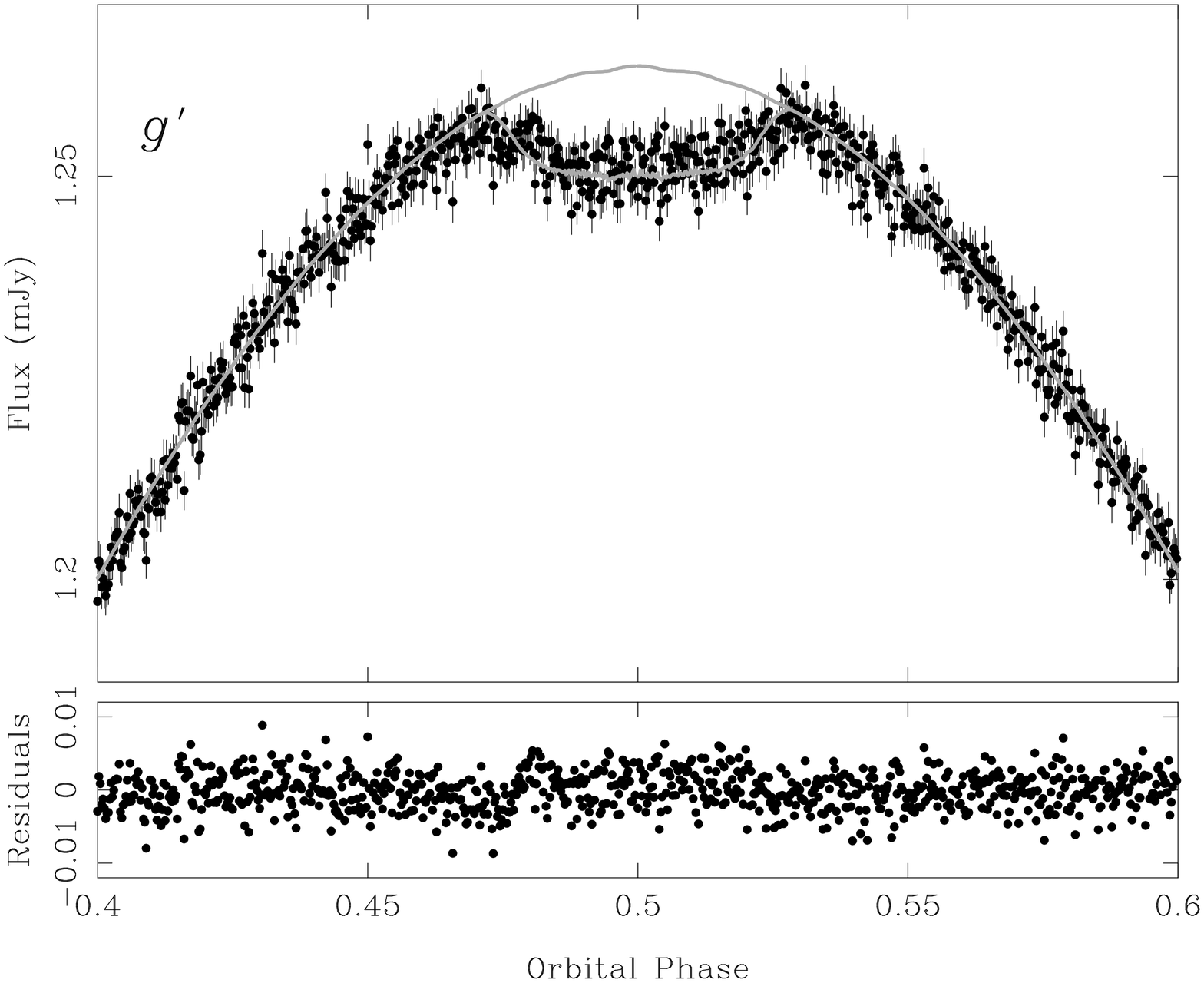}
 \includegraphics[width=0.3\textwidth]{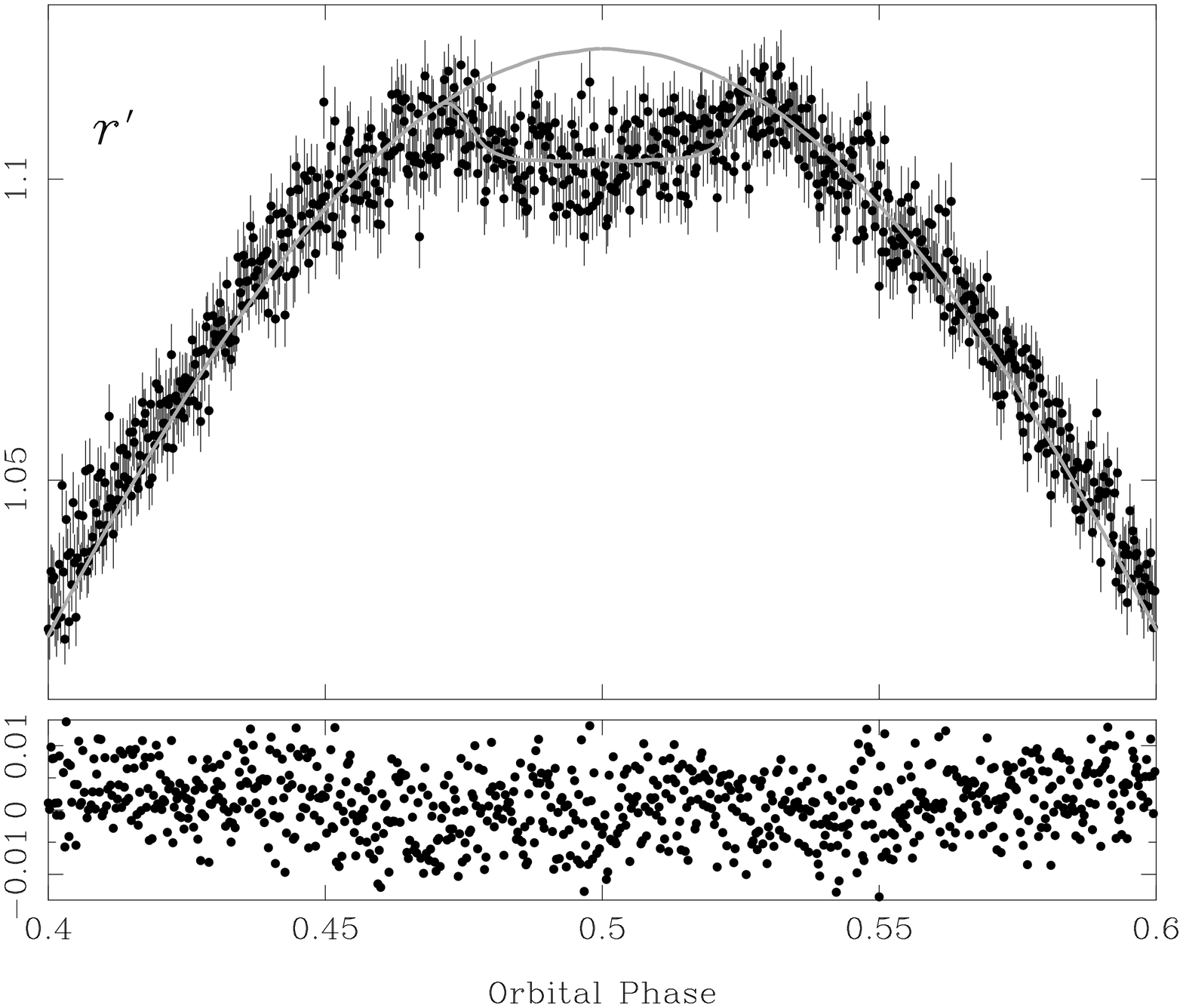}
 \includegraphics[width=0.3\textwidth]{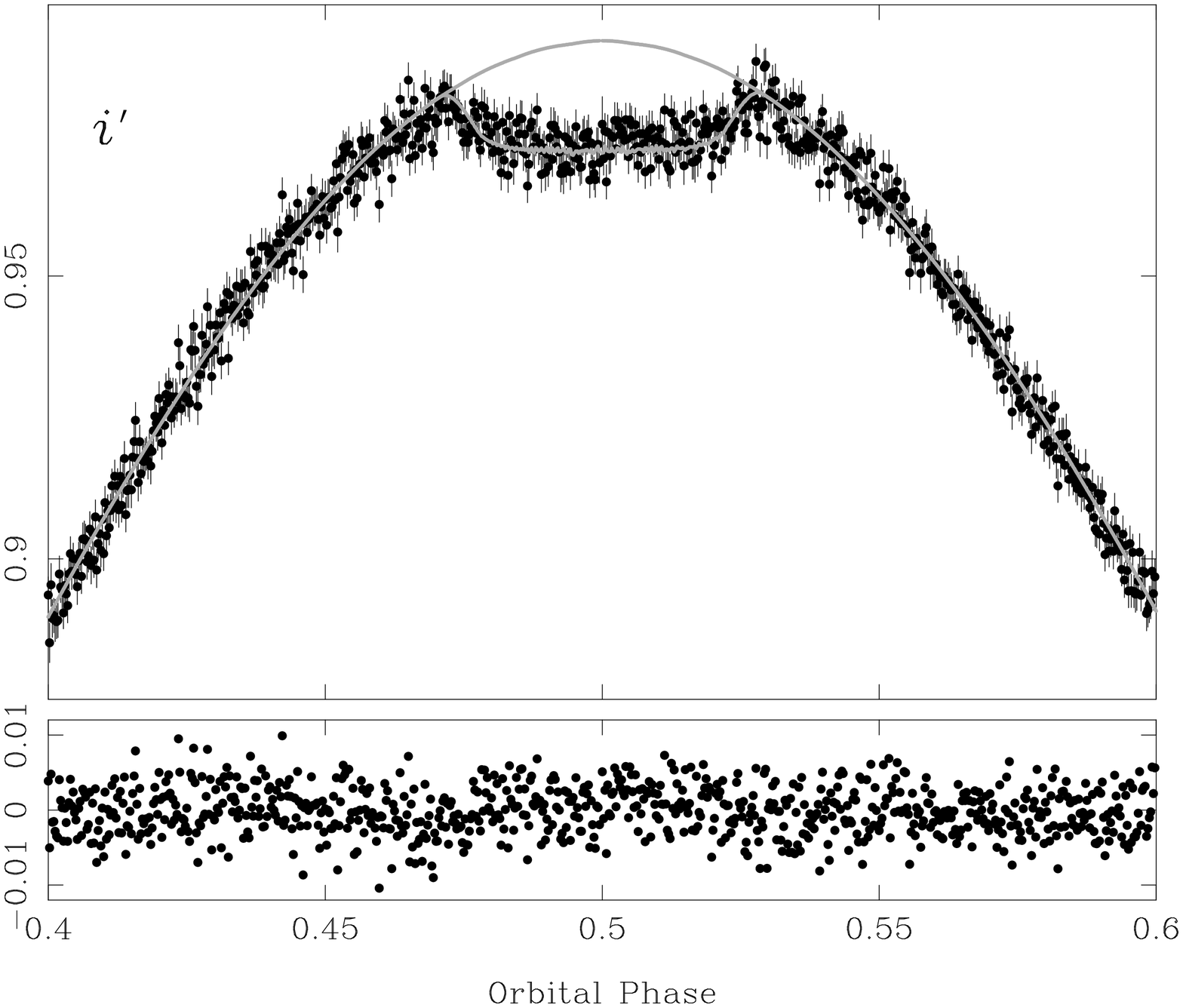}
 \caption{Model fits to the ULTRACAM light curves with residuals shown below. \emph{Top}: Full orbital phase. \emph{Centre}: Around the primary eclipse. \emph{Bottom}: Around the secondary eclipse. Finer binning was used around both the eclipses. The \emph{r'} light curves are slightly noisier around the eclipses because there is no VLT photometry for that filter. The secondary eclipse light curves are also shown with a model with the secondary eclipse turned off. Points around the primary and secondary eclipses were given twice the weighting of other points resulting in some residual effects seen in the residuals.}
 \label{lcurvefit}
\end{center}
\end{figure*}

Since phase-binned light curves were used, $T_{0}$ was set to zero but allowed to change while the period was kept fixed at 1. The primary and secondary eclipses are the most sensitive regions to the inclination and scaled radii. Hence, in order to determine the most accurate inclination and radii, the data around the two eclipses were given increased weighting in the fit (points with phases between 0.45 and 0.55 and between 0.95 and 0.05 were given twice the weighting of other points). We obtained \emph{z'} photometry for only one night and it was of fairly poor quality hence no model was fitted to it. The best fit parameters and their statistical errors are displayed in Table~\ref{mcmcfit} along with the linear limb darkening coefficients used for the white dwarf. Figure~\ref{lcurvefit} shows the fits to various light curves at different phases. For the secondary eclipse light curves, the same model is over-plotted but with the secondary eclipse turned off, demonstrating the high inclination of this system. The average $\chi^2$, per degree of freedom, for the fits was 1.7 for the \emph{g'}, \emph{r'} and \emph{i'} light curves and 2.1 for the \emph{u'} light curve. The MCMC chains showed no variation beyond that expected from statistical variance and the probability distributions are symmetrical and roughly Gaussian.

The errors in Table~\ref{mcmcfit} were scaled to give a reduced $\chi^2 = 1$. The inclination was determined by taking a weighted average and is found to be $89.6^{\circ} \pm 0.2^{\circ}$. This inclination is much higher than the $84.6^{\circ}$ determined by \citet{Haefner04} but is consistent with the inferred inclination of $\sim 88^{\circ}$ from \citet{Brinkworth06}. The scaled radius of the white dwarf is $R_\mathrm{WD}/a = 0.0226 \pm 0.0001$ and the scaled radius of the secondary star is $R_\mathrm{sec}/a = 0.165 \pm 0.001$. Given our black body assumption, $T_\mathrm{sec}$ does not represent the true temperature of the secondary star, it is effectively just a scaling factor. An interesting trend is seen in the limb darkening coefficients for the secondary star, which are all negative (limb brightening), the amount of limb brightening decreases with increasing wavelength. This is presumably the result of seeing to different depths at different wavelengths. 

Although the ULTRACAM \emph{z'} photometry was of fairly poor quality (owing to conditions), it was good enough to measure the magnitude of the secondary star during the primary eclipse. A zeroth-order polynomial was fit to the \emph{r'}, \emph{i'} and \emph{z'} filter light curves during the primary eclipse. The measured magnitudes were: \emph{r'}$ = 21.8 \pm 0.1$, \emph{i'}$ = 20.4 \pm 0.1$ and \emph{z'}$ = 19.6 \pm 0.1$, which gives colours of (\emph{r'-i'})$_\mathrm{sec} = 1.4 \pm 0.1$ and (\emph{i'-z'})$_\mathrm{sec} = 0.8 \pm 0.1$ which corresponds to a spectral type of M$4\pm0.5$ \citep{West05}. This is consistent with the results of \citet{Haefner04} who fitted the spectral features of the secondary star taken during the primary eclipse to determine a spectral type of M$4.75 \pm 0.25$.

\subsection{Heating of the Secondary Star}
One can make an estimate of the heating effect by comparing the intrinsic luminosity of the secondary star to that received from the white dwarf. We use the mass-luminosity relation from \citet{Scalo07} determined by fitting a polynomial to the luminosities and binary star masses compiled by \citet{Hillenbrand04} to determine the luminosity of the secondary star as $1.4 \times 10^{-3}$ L$_{\sun}$. The luminosity of the white dwarf was calculated using $L_\mathrm{WD}=4 \pi R^2 \sigma T^4$. Using the radius derived in Section 4.4 and the temperature from  \citet{Haefner04} gives the luminosity of the white dwarf as 4.2 L$_{\sun}$. Using the scaled radius of the secondary star from Table~\ref{mcmcfit} translates to the secondary star being hit by over 20 times its own luminosity. Despite this, the colours (hence spectral type) of the unirradiated side are in agreement with the derived mass \citep{Baraffe96} (see Section 4.4 for the mass derivation) showing that this extreme heating effect on one hemisphere of the secondary star has no measurable effect on the unirradiated hemisphere.

\begin{table}
 \centering
  \caption{Distance measurements from each of the ULTRACAM light curves. The absolute magnitudes for the white dwarf in NN Ser were obtained from \citet{Holberg06} with an error of $\pm 0.1$ magnitudes.}
  \label{distance}
  \begin{tabular}{@{}lcccc@{}}
  \hline
 Filter & Absolute & Measured & Extinction & Distance \\
        & Magnitude& Magnitude& (mags)     & (pc)    \\
 \hline 
 \emph{u'} & 7.264 & 15.992 $\pm$ 0.006 & 0.258 $\pm$ 0.258 & 494 $\pm$ 63 \\
 \emph{g'} & 7.740 & 16.427 $\pm$ 0.002 & 0.190 $\pm$ 0.190 & 501 $\pm$ 49 \\
 \emph{r'} & 8.279 & 16.931 $\pm$ 0.004 & 0.138 $\pm$ 0.138 & 505 $\pm$ 40 \\
 \emph{i'} & 8.666 & 17.309 $\pm$ 0.004 & 0.104 $\pm$ 0.104 & 510 $\pm$ 34 \\
 \emph{z'} & 9.025 & 17.71 $\pm$ 0.01   & 0.074 $\pm$ 0.074 & 527 $\pm$ 30 \\
 \hline
\end{tabular}
\end{table}

\subsection{Distance to NN Ser}
Absolute magnitudes for the white dwarf in NN Ser were calculated using a model from \citet{Holberg06} for a DA white dwarf of mass 0.527M$_{\sun}$, $\log{g} = 7.5$ and a temperature of 60,000K which most closely matched the parameters found for NN Ser. We give an uncertainty of $\pm0.1$ magnitudes for the absolute magnitudes based on the uncertainty in temperature from \citet{Haefner04} and its effect on the models of \citet{Holberg06}. The magnitudes of the white dwarf in NN Ser were calculated by fitting a zeroth-order polynomial to the flat regions either side of the primary eclipse with a correction made for the flux of the secondary star. Using the reddening value of E(B-V)$ = 0.05 \pm 0.05$ from \citet{Wood91} we correct the apparent magnitudes using the conversion of \citet{Schlegel98}. From these a distance was calculated for each of the ULTRACAM filters. Table~\ref{distance} lists the distances calculated in each of these filters. Using these values gives a distance to NN Ser of $512 \pm 43$ pc consistent with the result of \citet{Haefner04} of $500 \pm 35$ pc.

\begin{figure}
\begin{center}
 \includegraphics[width=\columnwidth]{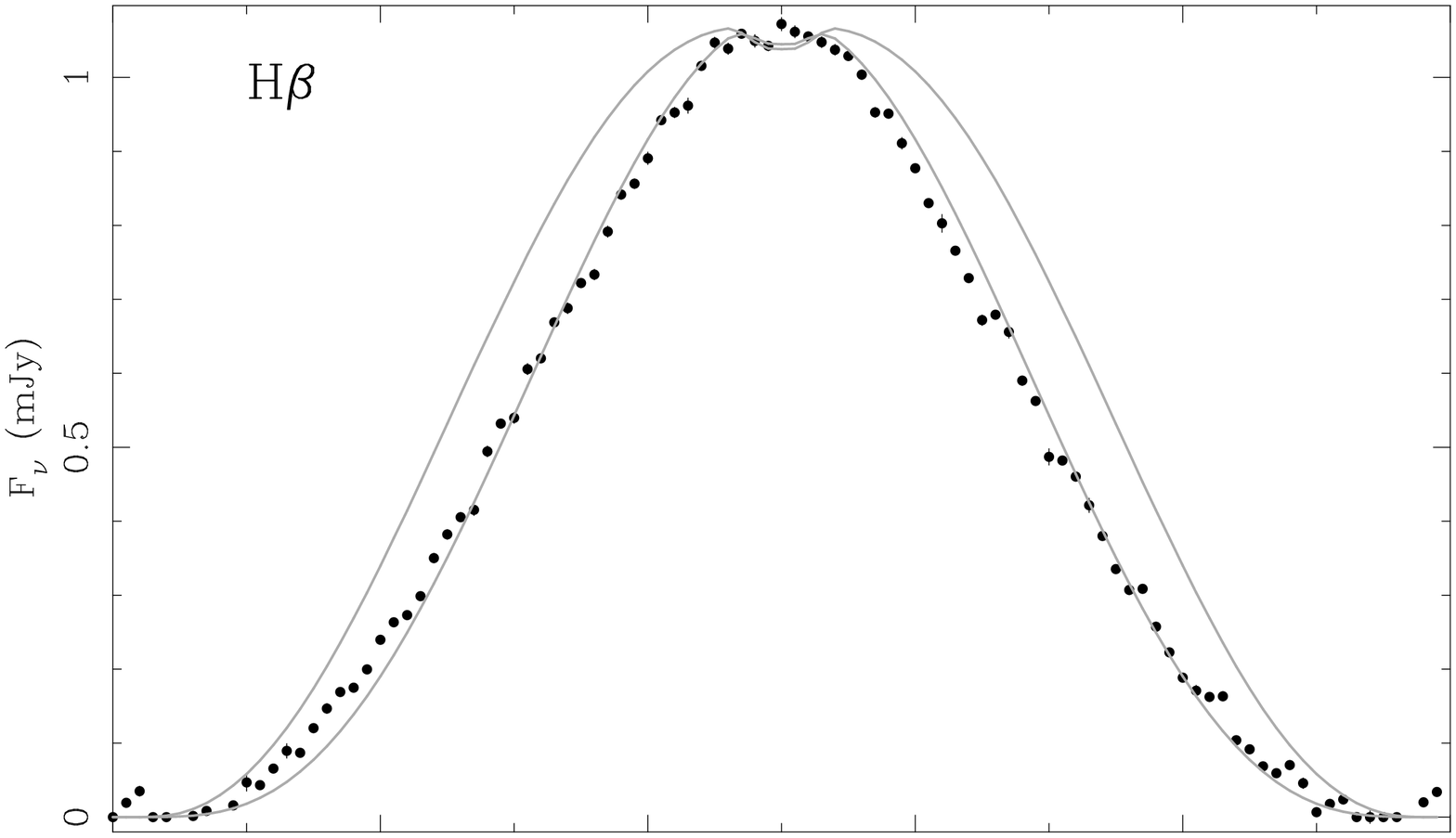}
 \includegraphics[width=\columnwidth]{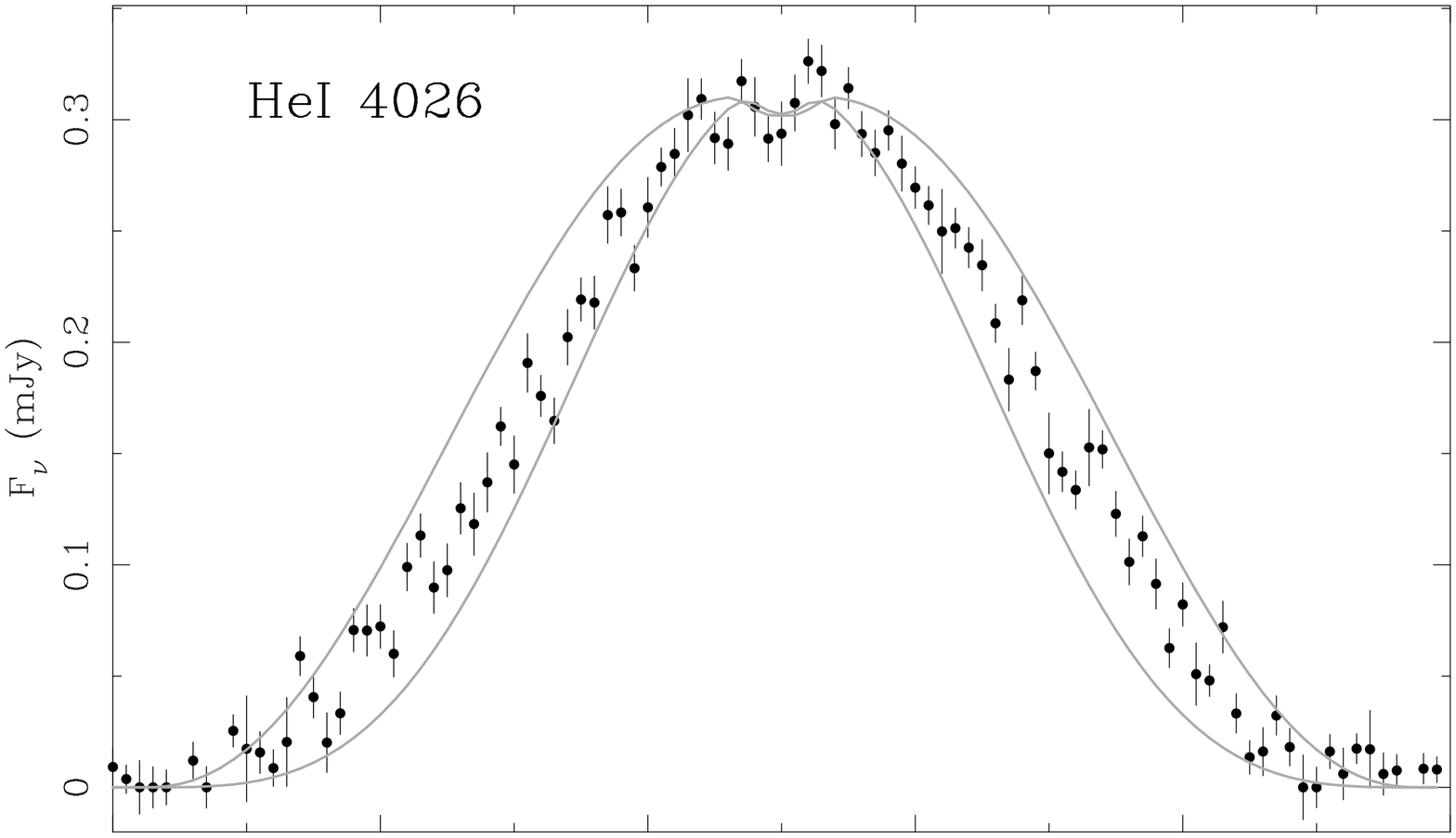}
 \includegraphics[width=\columnwidth]{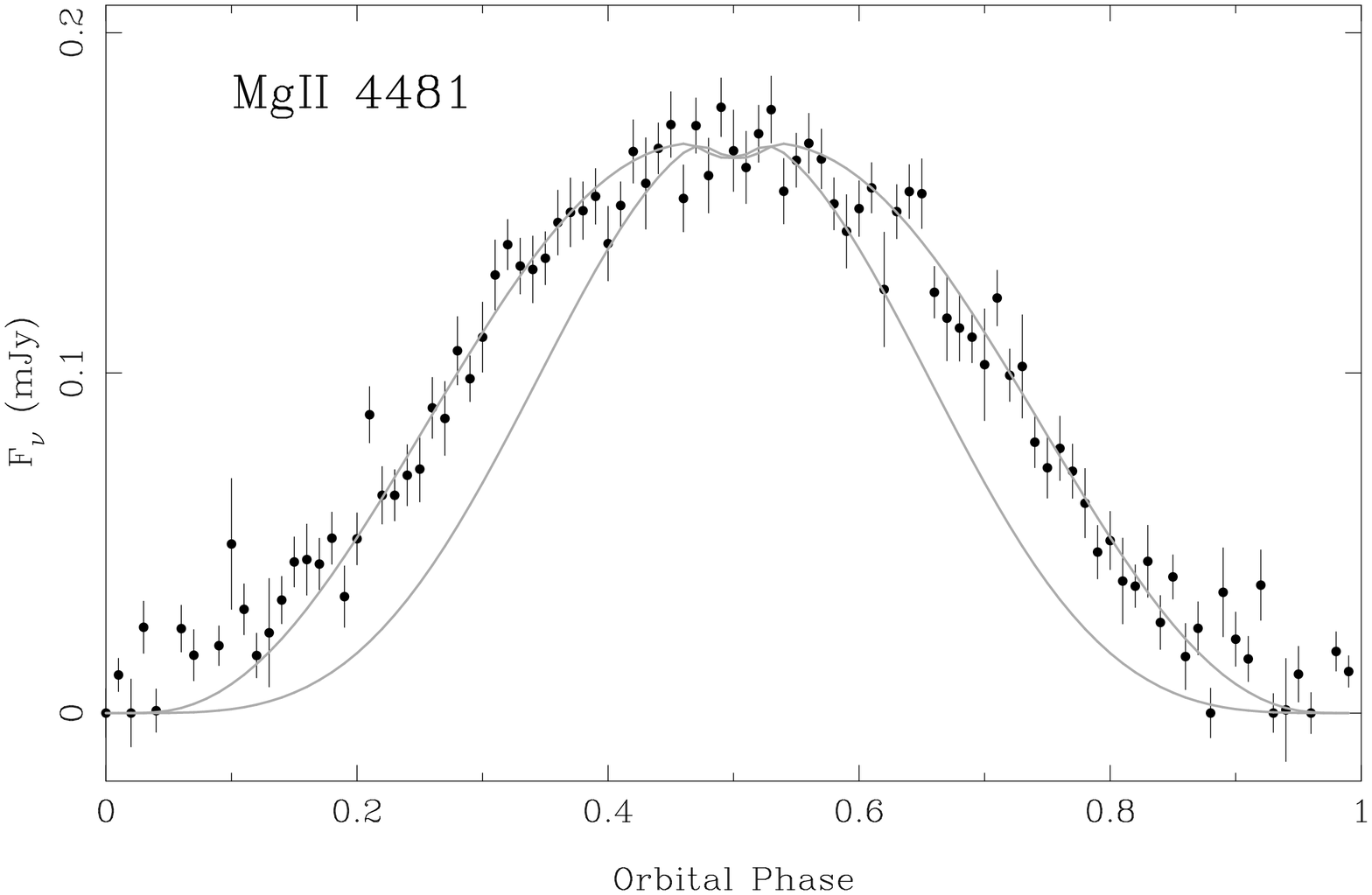}
 \caption{Light curves of various lines over-plotted with an optically thick (narrower) and thin (wider) model with the same measured $K_\mathrm{sec}$ as the line. The model light curves are scaled to match the flux level of the lines around phase 0.5. The H$\beta$ line is optically thick, the MgII line is optically thin and the HeI line is somewhere between thick and thin.}
 \label{lcurvemag}
\end{center}
\end{figure}

The galactic latitude of NN Ser is $45.3^{\circ}$ which, combined with the derived distance, gives NN Ser a galactic scale height of $364 \pm 31$ pc. The proper motion of NN Ser was retrieved from the US Naval Observatory (USNO) Image and Catalogue Archive. The archive values are $\mu_\mathrm{RA}=-0.020 \pm 0.003$ and $\mu_\mathrm{DEC}=-0.056 \pm 0.004$ arcsec / yr. At the derived distance this corresponds to a transverse velocity for NN Ser of $160 \pm 14$  km$\,$s$^{-1}$.

\subsection{$K_\mathrm{sec}$ correction}
The emission lines seen in the UVES spectra are the result of reprocessed light from the surface of the secondary star facing the white dwarf. Hence, their radial velocity amplitude represents a lower limit to the true centre of mass radial velocity amplitude. For accurate mass determinations the centre of mass radial velocity amplitude is required thus we need to determine the deviation between the reprocessed light centre and the centre of mass for the secondary star. We do this by computing model line profiles from the irradiated face.

We use the inclination and radii determined from the light curve model and account for Roche distortion of the secondary star and the secondary eclipse. All other parameters are set to match the UVES spectra: the line profiles were calculated for phases matching the UVES spectra phases, exposure lengths were set to the same as the UVES blue spectra exposures and sampled in velocity to match the spectra. The profiles are convolved with a Gaussian function, representing the resolution of the spectrograph. In order to match the UVES spectroscopy as closely as possible, the resolution of the spectrograph was calculated by looking at the arc calibration spectra. The intrinsic linewidth of these lines is assumed to be negligible hence the measured linewidth gives an indication of the instrumental resolution, which we found to be 5 km$\,$s$^{-1}$ (FWHM) for the UVES blue chip.

\begin{table}
 \centering
  \caption{Measured and corrected values of $K_\mathrm{sec}$  with the best fit model line profiles parameters, for several lines in the UVES blue spectra.}
  \label{irradprof}
  \begin{tabular}{@{}lccccc@{}}
  \hline
Line & $\tau_0$ &  $\beta$ & $K_\mathrm{sec,meas}$ & $K_\mathrm{sec,corr}$ & $q$ \\
 & & & km$\,$s$^{-1}$ & km$\,$s$^{-1}$ & \\
 \hline 
 HeI 3820   & $2$    & $-0.75$ & $252.2 \pm 1.9$ & $296.7 \pm 1.9$ & 0.210(6) \\
 HeI 4026   & $1$    & $-3$    & $257.7 \pm 1.7$ & $300.1 \pm 1.7$ & 0.208(6) \\
 HeI 4388   & $1$    & $-10$   & $249.9 \pm 2.0$ & $300.2 \pm 2.0$ & 0.208(6) \\
 HeI 4472   & $1$    & $-1.5$  & $263.1 \pm 1.8$ & $305.2 \pm 1.8$ & 0.204(6) \\
 HeI 4922   & $1$    & $-0.5$  & $255.6 \pm 1.8$ & $296.9 \pm 1.8$ & 0.210(6) \\
 H$\beta$   & $100$  & $-1.5$  & $271.1 \pm 1.9$ & $307.6 \pm 1.9$ & 0.203(6) \\
 H$\gamma$  & $100$  & $-1.5$  & $265.1 \pm 1.9$ & $302.9 \pm 1.9$ & 0.206(6) \\
 H$\delta$  & $50$   & $-1.5$  & $265.0 \pm 1.9$ & $298.1 \pm 1.9$ & 0.209(6) \\
 H$\epsilon$& $5$    & $-1.25$ & $263.2 \pm 1.9$ & $304.4 \pm 1.9$ & 0.205(6) \\
 H8         & $5$    & $-1$    & $262.5 \pm 1.9$ & $303.2 \pm 1.9$ & 0.206(6) \\
 H9         & $2$    & $-0.5$  & $258.1 \pm 1.9$ & $297.4 \pm 1.9$ & 0.210(6) \\
 H10        & $2$    & $-0.75$ & $257.2 \pm 2.1$ & $299.9 \pm 2.1$ & 0.208(6) \\
 H11        & $1$    & $-2$    & $259.3 \pm 2.1$ & $301.2 \pm 2.1$ & 0.207(6) \\
 MgII 4481  & $0.005$& $-1.25$ & $252.8 \pm 1.8$ & $298.7 \pm 1.8$ & 0.209(6) \\
 \hline
\end{tabular}
\end{table}

As mentioned previously, the measured radial velocity amplitude varies for each line. We believe that this scatter is the result of differences in optical depths of the lines, which will affect the angular distribution of line flux from any given point on the star resulting in a range of observed radial velocity amplitudes. Hence, the model is required to produce line profiles over a continuous range of optical depths (see Appendix for details of the model).

The radiation from the secondary star is modelled as a slab of constant optical depth $(\tau_0)$ and the source function changes exponentially with depth, the factor that determines how this changes is $\beta$ (i.e. the source function  changes with vertical optical depth ($\tau$) as e$^{\beta\tau}$), this allows one to have a continuous transition from optically thin to thick and to have limb darkening or brightening. We keep $K_\mathrm{WD}$ fixed at the measured value of $62.3$ km$\,$s$^{-1}$, and just change $K_\mathrm{sec}$. We measured the radial velocity amplitude of the resulting line profiles in the same way as for the emission lines in the UVES spectra. Initially, $K_\mathrm{sec}$ was set to give a measured radial velocity amplitude of $252.8$ km$\,$s$^{-1}$ (the measured radial velocity amplitude of the MgII 4482\AA~ line) and the values of the total vertical optical depth and source function exponential factor were allowed to vary. The light curves produced were fitted to the MgII 4481\AA~ line light curve using least squares fitting to determine the optimal values for $\tau_0$ and $\beta$. This was repeated for several lines in the UVES blue spectra adjusting $K_\mathrm{sec}$ to produce the measured radial velocity amplitude for that line. Figure~\ref{lcurvemag} shows the light curve for three different lines over plotted with an optically thick and optically thin model. For the H$\beta$ line, the emission is optically thick, the MgII line is optically thin and the HeI line lies somewhere between these two extremes (the white dwarf component was subtracted from all the light curves). Table~\ref{irradprof} lists the best fit values for $\tau_0$ and $\beta$ for each line and the measured and corrected $K_\mathrm{sec}$ and $q$ values. 

\begin{figure}
\begin{center}
 \includegraphics[width=\columnwidth]{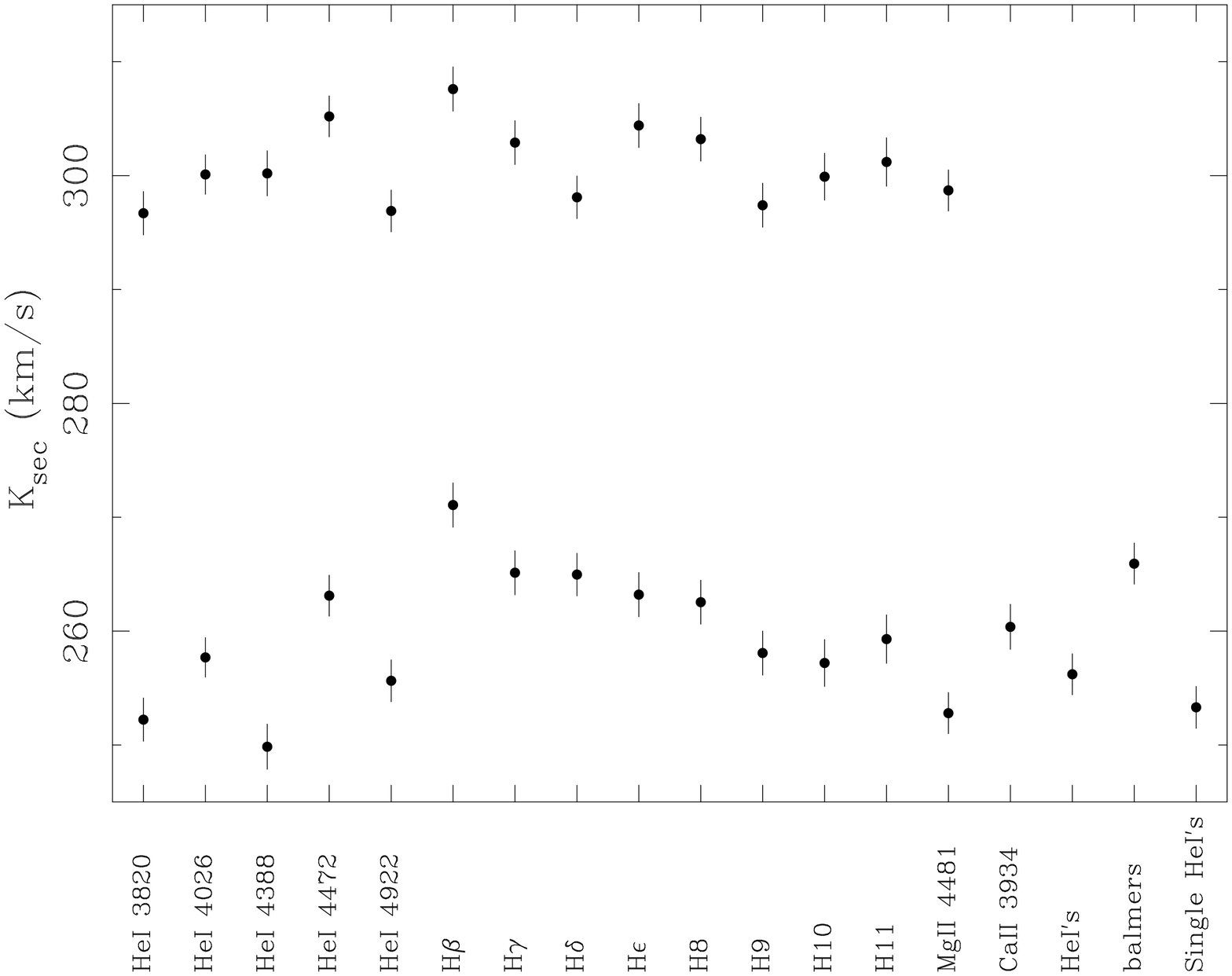}
 \caption{\emph{Bottom}: Measured $K_\mathrm{sec}$ for several lines in the UVES blue spectra from Gaussian fitting and their corrected values (\emph{top}). The CaII line is contaminated by interstellar absorption which was fitted as an additional Gaussian component and was not corrected. The fit to all the Balmer lines simultaneously is the point labelled `balmers'. The point labelled `HeI's' is a fit to all the visible HeI lines whilst the point labelled `single HeI's' is a fit to only the HeI lines which have a single component (i.e. the intensity is not shared by several lines).}
 \label{kcorr}
\end{center}
\end{figure}

The MgII 4481\AA~ line appears to be the closest to the optically thin model. As such it probably provides the most accurate correction since, in the optically thin case, the angular distribution of the line flux from any given point on the star is even, removing any dependence upon this. Even so, all the corrected values are consistent to within a few km$\,$s$^{-1}$. The first few Balmer lines (H$\beta$ to H$\delta$) are optically thick but as the series progresses, the lines become more optically thin. There also appears to be a small increase in the value of $\beta$ throughout the series (with the exception of H10 and H11). The helium lines appear to be somewhere between optically thick and thin. 

Figure~\ref{kcorr} shows the measured values of $K_\mathrm{sec}$ for several lines from Gaussian fitting in the UVES blue spectra and their corrected values of $K_\mathrm{sec}$. The spread in values is reduced and the corrected values give a radial velocity amplitude for the secondary star of $K_\mathrm{sec} = 301$ km$\,$s$^{-1}$. The statistical uncertainty is $1$ km$\,$s$^{-1}$, however, we believe the error in the correction is dominated by systematic effects from the model and fitting process, for which we estimate an error of $3$ km$\,$s$^{-1}$ for the radial velocity amplitude of the secondary star. Some simple considerations can give an idea of the largest possible error we might have made in correcting the $K$ values. The distance from the centre of the mass to the sub-stellar point in terms of velocity is given by $\left(K_\mathrm{WD}+K_\mathrm{sec}\right) R_\mathrm{sec}/a = 60 \,\mathrm{km}\,\mathrm{s}^{-1}$; this is maximum possible correction. A lower limit comes from assuming the emission to be uniform over the irradiated face of the secondary. Then the centre of light is $0.42$ of the way from the centre of mass to the sub-stellar point \citep{Wade88}, leading to a lower limit on the correction of $24\,\mathrm{km}\,\mathrm{s}^{-1}$. The corrections from our model range from 33 to $46\,\mathrm{km}\,\mathrm{s}^{-1}$, in the middle of these extremes. 

\begin{table}
 \centering
  \caption{System parameters. (1) this paper; (2) \citet{Brinkworth06}; (3) \citet{Haefner04}. WD: white dwarf, Sec: secondary star.}
  \label{sys_para}
  \begin{tabular}{@{}lll@{}}
  \hline
  Parameter              &  Value                          & Ref. \\
 \hline
  Period (days)          &  0.13008017141(17)              & (2) \\
  Inclination            &  89.6 $\pm$ $0.2^{\circ}$        & (1) \\
  Binary separation      &  0.934 $\pm$ 0.009 R$_{\sun}$    & (1) \\
  Mass ratio             &  0.207 $\pm$ 0.006              & (1) \\
  WD mass                &  0.535 $\pm$ 0.012 M$_{\sun}$    & (1) \\
  Sec mass               &  0.111 $\pm$ 0.004 M$_{\sun}$    & (1) \\
  WD radius              &  0.0211 $\pm$ 0.0002 R$_{\sun}$  & (1) \\
  Sec radius polar       &  0.147 $\pm$ 0.002 R$_{\sun}$    & (1) \\
  Sec radius sub-stellar &  0.154 $\pm$ 0.002 R$_{\sun}$    & (1) \\
  Sec radius backside    &  0.153 $\pm$ 0.002 R$_{\sun}$    & (1) \\
  Sec radius side        &  0.149 $\pm$ 0.002 R$_{\sun}$    & (1) \\
  Sec radius volume-averaged &  0.149 $\pm$ 0.002 R$_{\sun}$    & (1) \\
  WD $\log{g}$           &  7.47 $\pm$ 0.01                & (1) \\
  WD temperature         &  57000 $\pm$ 3000 K             & (3) \\
  $K_\mathrm{WD}$         &  62.3 $\pm$ 1.9 km$\,$s$^{-1}$   & (1) \\
  $K_\mathrm{sec}$        &  301 $\pm$ 3 km$\,$s$^{-1}$      & (1) \\
  WD  grav. redshift     &  10.5 $\pm$ 2.7 km$\,$s$^{-1}$   & (1) \\
  WD type                &  DAO1                           & (3) \\
  Sec spectral type      &  M$4\pm0.5$                       & (1) \\
  Distance               &  512 $\pm$ 43 pc                & (1) \\
  WD hydrogen layer fractional mass & $10^{-4}$            & (1) \\
 \hline
\end{tabular}
\end{table}

\begin{figure}
\begin{center}
 \includegraphics[width=\columnwidth]{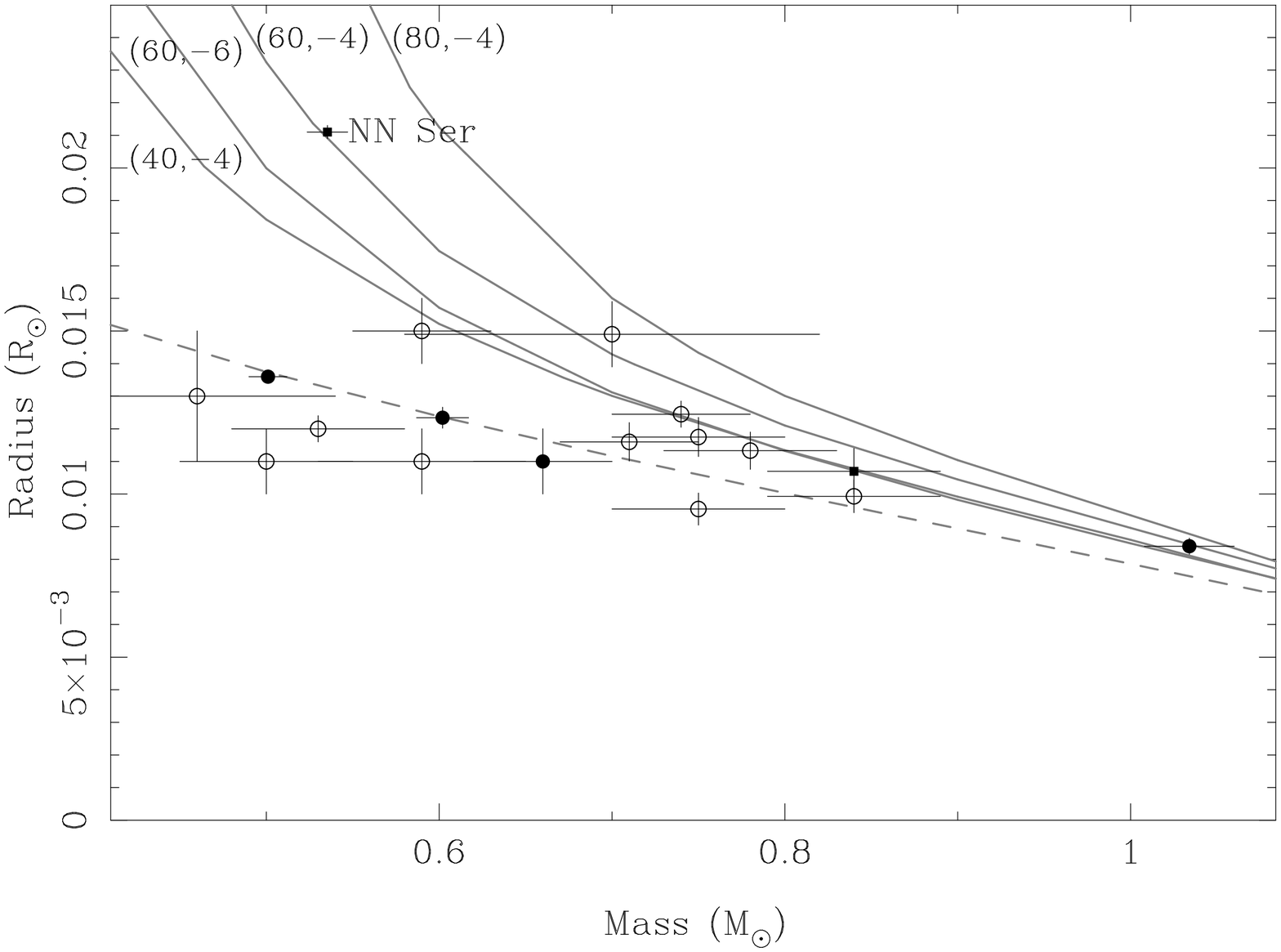}
 \caption{Mass-radius plot for white dwarfs measured independent of any mass-radius relations. Data from \citet{Provencal98}, \citet{Provencal02} and \citet{Casewell09} are plotted. The filled circles are visual binaries and the open circles are common proper-motion systems. The solid lines correspond to different carbon-oxygen core pure hydrogen atmosphere models. The first number is the temperature, in thousands of degrees, the second number is the hydrogen layer thickness (i.e. lines labelled -4 have a thickness of M$_\mathrm{H}/\mathrm{M}_\mathrm{WD} = 10^{-4}$) from \citet{Holberg06} and \citet{Benvenuto99}. The dashed line is the zero-temperature mass-radius relation of Eggleton from \citet{Verbunt88}.}
 \label{wdmr}
\end{center}
\end{figure}

From the corrected value and $K_\mathrm{WD} = 62.3 \pm 1.9$ km$\,$s$^{-1}$ we determine the mass ratio as $q = M_\mathrm{sec}/M_\mathrm{WD} = 0.207 \pm 0.006$. Using this, the mass of the white dwarf is then determined using Kepler's third law
\begin{eqnarray}
\frac{P K_\mathrm{sec}^3}{2 \pi G} = \frac{M_\mathrm{WD} \sin^3{i}}{(1+q)^2},
\end{eqnarray}
using the period $(P)$ from \citet{Brinkworth06}. This gives a value of $M_\mathrm{WD} = 0.535 \pm 0.012$ M$_{\sun}$. The mass ratio then gives the mass of the secondary star as $M_\mathrm{sec} = 0.111 \pm 0.004$ M$_{\sun}$. Knowing the masses, the orbital separation is found using
\begin{eqnarray}
a^3 = \frac{G (M_\mathrm{WD} + M_\mathrm{sec}) P^2}{4 \pi^2},
\end{eqnarray}
which gives a value of $a = 0.934 \pm 0.009$ R$_{\sun}$. Using this gives the radii of the two stars as $R_\mathrm{WD} = 0.0211 \pm 0.0002$ R$_{\sun}$ and $R_\mathrm{sec} = 0.154 \pm 0.002$ R$_{\sun}$. The radius of the secondary star in our model is measured from the centre of mass towards the white dwarf and hence is larger than the average radius of the secondary star. The radius as measured towards the backside is $R_\mathrm{sec,back} = 0.153$ R$_{\sun}$, the polar radius is $R_\mathrm{sec,pol} = 0.147$ R$_{\sun}$ and perpendicular to these $R_\mathrm{sec,side} = 0.149$ R$_{\sun}$. The volume-averaged radius is $R_\mathrm{sec,av} = 0.149 \pm 0.002$ R$_{\sun}$.

The surface gravity of the white dwarf is given by
\begin{eqnarray}
g = \frac{G M_\mathrm{WD}}{R_\mathrm{WD}{}^2},
\end{eqnarray}
which gives a value of $\log{g} = 7.47 \pm 0.01$.  

\section{Discussion}

Figure~\ref{wdmr} shows the mass and radius of the white dwarf of NN Ser in comparison with other white dwarfs whose mass and radius have been measured independently of any mass-radius relation. The white dwarf of NN Ser has the largest measured radius of any white dwarf (not determined by a mass-radius relation) placing it in a new area of this plot; it is also one of the most precisely measured. NN Ser is also much hotter than the other white dwarfs in Figure~\ref{wdmr}, the mean temperature of the other white dwarfs is $\sim 14,200$K. Also plotted are several mass-radius relations for pure hydrogen atmosphere white dwarfs with different temperatures and hydrogen layer thicknesses from \citet{Holberg06} and \citet{Benvenuto99}. The earlier work from \citet{Haefner04} and the UVES spectrum of the white dwarf with a model white dwarf spectra over-plotted (Figure~\ref{wdfull}) strongly suggest a temperature for the white dwarf of close to 60,000K ($57000 \pm 3000$K). Assuming the models in Figure~\ref{wdmr} are correct, the white dwarf in NN Ser shows excellent agreement with having a `thick' hydrogen layer of fractional mass M$_\mathrm{H}/\mathrm{M}_\mathrm{WD} = 10^{-4}$. This is consistent with the inflated radius of the white dwarf due to its high temperature. 

Since visual binary systems and common proper-motion systems still rely on model atmosphere calculations to determine radii, the white dwarf in NN Ser is one of the first to have its mass and radius measured independently. \citet{Obrien01} determine the mass and radius of both components of the eclipsing PCEB V471 Tau independently however, since they did not detect a secondary eclipse, they had to rely on less direct methods to determine the radius of the secondary and the inclination. This demonstrates the value of eclipsing PCEBs for investigating the mass-radius relation for white dwarfs. 

\begin{figure}
\begin{center}
 \includegraphics[width=\columnwidth]{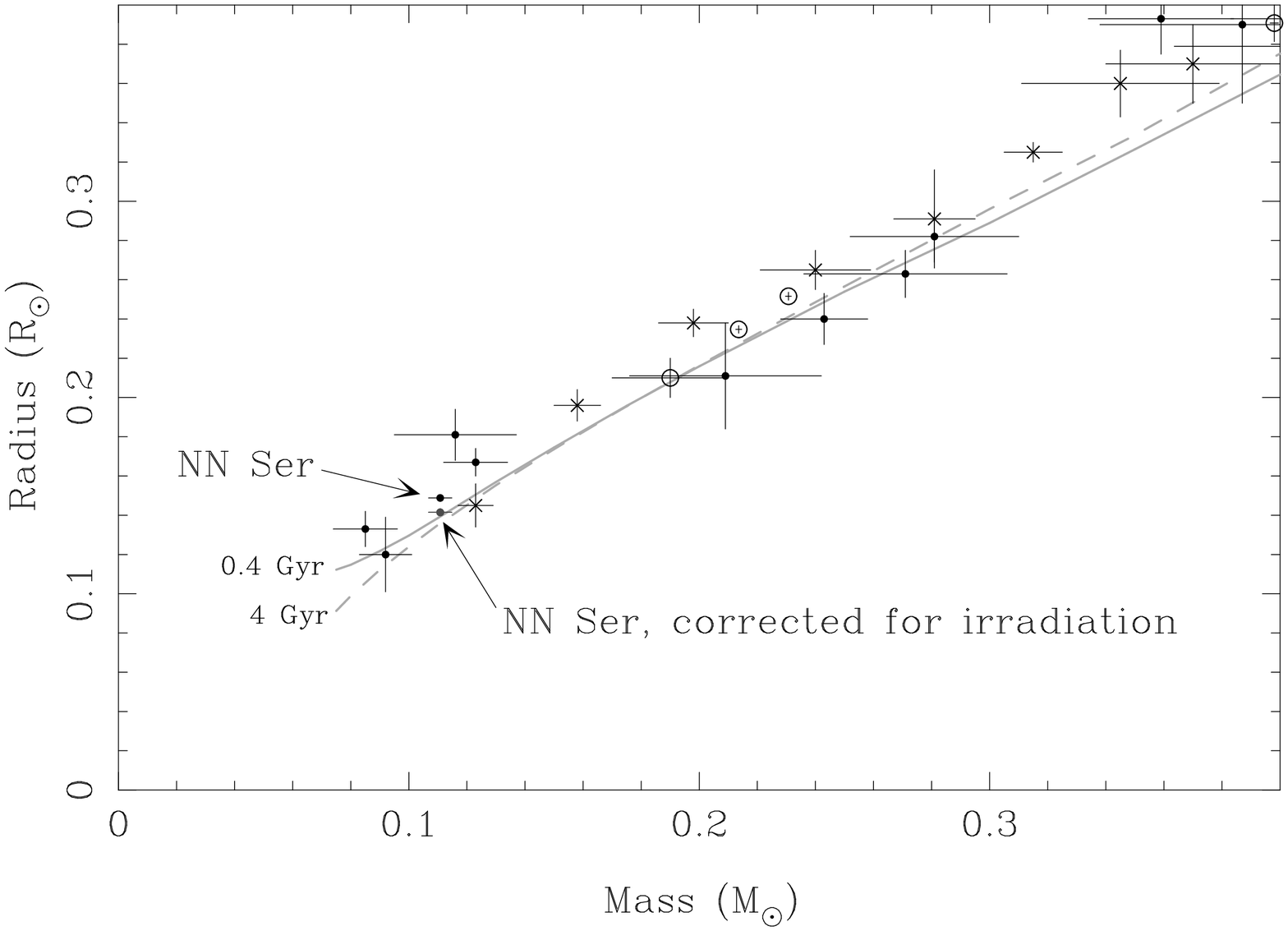}
 \caption{Mass-radius plot for low mass stars. Data from \citet{Lopez07} and \citet{Beatty07}. The filled circles are secondaries in binaries, the open circles are low mass binaries and the crosses are single stars. The solid line represents the theoretical isochrone model from \citet{Baraffe98}, for an age of 0.4 Gyr, solar metalicity, and mixing length $\alpha = 1.0$, the dotted line is the same but for an age of 4 Gyr. The position of the secondary star in NN Ser is also shown if there were no irradiation effects. The masses of the single stars were determined using mass-luminosity relations.}
 \label{rdmr}
\end{center}
\end{figure}

In addition, the mass and radius of the secondary star have been determined independently of any mass-radius relation. Since this is a low mass star it helps improve the statistics for these objects and our values are more precise than the majority of comparable measurements. Figure~\ref{rdmr} shows the position of the secondary star in NN Ser (using the volume-averaged radius) in relation to other low mass stars with masses and radii determined independently of any mass-radius relation (although the masses of the single stars were determined using mass-luminosity relations). For very low mass stars (M $\la 0.3$ M$_{\sun}$) the current errors on mass and radius measurements are so large that that one can argue the data are consistent with the low-mass models. However, the secondary star in NN Ser appears to be the first object with errors small enough to show an inconsistency with the models. The measured radius is 10\% larger than predicted by the model. However, irradiation increases the radius of the secondary star. For the measured radius of the secondary star in NN Ser, the work of \citet{Ritter00} and \citet{Hameury97} gives an inflation of 5.6\%. The un-irradiated radius is also shown in Figure~\ref{rdmr} and is consistent with the models. Hence the secondary star in NN Ser supports the theoretical mass-radius relation for very low mass stars. Potentially, an initial-final mass relation could be used to estimate the age of NN Ser, but since the system has passed through a common envelope phase its evolution may have been accelerated and the white dwarf may be less massive than a single white dwarf with the same progenitor mass leading to an overestimated age. In addition, the mass of the white dwarf in NN Ser is close to the mean white dwarf mass and the initial-final mass relation is flat in this region. This means a large range of progenitor masses are possible for the white dwarf and hence a large range in age, this means a reliable estimate of the age of NN Ser is not possible. In any case, the position of the un-irradiated secondary star is also consistent with a similarly large range of ages. The mass and radius of the secondary star support the argument of \citet{Brinkworth06} that it is not capable of generating enough energy to drive period change via Applegate's mechanism \citep{Applegate92}. 

The system parameters determined for NN Ser are summarised in Table~\ref{sys_para}. Using the UVES spectra, the gravitational redshift of the white dwarf was found to be $10.5 \pm 2.7$ km$\,$s$^{-1}$. Using the measured mass and radius from Table~\ref{sys_para} gives a redshift of $16.1$ km$\,$s$^{-1}$, correcting for the redshift of the secondary star (0.5 km$\,$s$^{-1}$), the difference in transverse Doppler shifts (0.15 km$\,$s$^{-1}$) and the potential at the secondary star owing to the white dwarf (0.3 km$\,$s$^{-1}$) gives a value of $15.2 \pm 0.5$ km$\,$s$^{-1}$ which is consistent with the measured value to $\sim 2$ sigma, although in this case, the inflated radius of the white dwarf weakens the constraints of the gravitational redshift. 

\section{Conclusions}
We have measured precise masses and radii of the white dwarf and M dwarf components of the post common envelope binary NN Serpentis using UVES spectroscopy and ULTRACAM photometry.

Using the HeII 4686\AA~ absorption line from the white dwarf we determined the radial velocity amplitude of the white dwarf directly from the spectra. Using a number of emission lines in the UVES spectra originating from the heated face of the secondary star, we were able to correct the radial velocity amplitude of the secondary star from the heated face to the centre of mass of the secondary star itself.

From analysis of ULTRACAM light curves we determine a system inclination of $89.6^{\circ} \pm 0.2^{\circ}$, higher than the value of $84.6^{\circ} \pm 1.1^{\circ}$ found by \citet{Haefner04} which, along with our direct determination of $K_\mathrm{WD}$, leads to a lower mass ratio than previously derived. The radius of the white dwarf is found to be $R_\mathrm{WD} = 0.0211 \pm 0.0002$ R$_{\sun}$, larger than in previous studies but, given its temperature, is consistent with its derived mass of $M_\mathrm{WD} = 0.535 \pm 0.012$ M$_{\sun}$. The mass and radius of the white dwarf show excellent agreement with a hot carbon-oxygen core white dwarf with a `thick' hydrogen layer of fractional mass M$_\mathrm{H}/\mathrm{M}_\mathrm{WD} = 10^{-4}$. 

The mass of the secondary star is found to be $M_\mathrm{sec} = 0.111 \pm 0.004$ M$_{\sun}$ with a volume-averaged radius of $R_\mathrm{sec} = 0.149 \pm 0.002$ R$_{\sun}$, which is smaller than previously determined. The radius of the secondary star is consistent with models if a $\sim 6$\% correction is made for the irradiation it receives from the white dwarf. 

The ULTRACAM photometry also provided colours for the secondary star and thus a spectral type. This was consistent with the derived mass showing that, despite being irradiated by over 20 times its own luminosity, there is very little backside heating, although infrared data are needed to determine this more accurately. Finally, using model white dwarf data we determine a distance to NN Ser of 512 $\pm$ 43 pc, consistent with previous studies.

\section*{Acknowledgements}
We thank the referee, M Burleigh, for his useful comments and suggestions. TRM, CMC and BTG acknowledge support from the Science and Technology Facilities Council (STFC) grant number ST/F002599/1. SPL acknowledges the support of an RCUK Fellowship. ULTRACAM, VSD and SPL are supported by STFC grants ST/G003092/1 and PP/E001777/1. The results presented in this paper are based on observations collected at the European Southern Observatory (La Silla) under programme ID 073.D-0633 and with the William Herschel Telescope operated on the island of La Palma by the Isaac Newton Group in the Spanish Observatorio del Roque de los Muchachos of the Institutions de Astrofisica de Canarias. We used SIMBAD, maintained by the Centre de Donn\'{e}es astronomiques de Strasbourg, and the National Aeronautics and Space Administration (NASA) Astrophysics Data System. This research has made use of the USNOFS Image and Catalogue Archive operated by the United States Naval Observatory, Flagstaff Station and the National Institute of Standards and Technology (NIST) Atomic Spectra Database (version 3.1.5). STScI is operated by the Association of Universities for Research in Astronomy inc.

\bibliographystyle{mn2e}
\bibliography{nnser}

\appendix

\section{Modelling of the irradiation lines}

The radial velocity semi-amplitudes we measure for the emission lines in NN Ser reflect the distance from the centre of mass of the binary of the irradiated face of the secondary star. To obtain the semi-amplitude of the secondary star, $K_\mathrm{sec}$, we need to correct for the distance from the flux-weighted mean of the irradiated flux to the centre of mass of the secondary. To do this we need to know the size of the secondary star, which we know accurately from photometry, but also the distribution of flux. We adopted an empirical modelling approach which is described in this section.

To model the irradiated flux we modelled the surface of the secondary star as a series of small elements, allowing for the (small) distortion from tidal and centrifugal forces. The intensity of irradiated flux from each point was set to be linearly proportional to the incident flux from the white dwarf allowing for inverse square law dilution and incident angle. Our ultimate goal was to simulate the line profiles so that we could measure radial-velocities from them to allow us to adjust $K_\mathrm{sec}$ until we matched the observed semi-amplitude. As Table~\ref{irradprof} and Figure~\ref{kcorr} show however, the observed semi-amplitudes varied from line to line over a range of $20\,\mathrm{km}\,\mathrm{s}^{-1}$. We believe that this reflects differences in optical depths in the lines, which will affect the angular distribution of line flux from any given point on the star. For instance, if the flux is preferentially beamed perpendicular to the stellar surface, then at the quadrature phases, we will see the limb of the irradiated region more prominently compared to the region of maximum irradiation than we would otherwise. This will lead to a higher observed semi-amplitude. To allow for such effects we devised a simple model of the line emitting region, which we now describe.

\subsection{Optical depth model}

We wanted to be able to model optically thin and optically thick emitting regions within one model so that there was a continuous transition from one to the other. To do so we assumed a simple model in which the line emitting region at any point on the secondary behaves as if it had a total vertical optical depth $\tau_0$, and a source function given by an exponential function of vertical optical depth, $\tau$,
\[ S(\tau) \propto e^{\beta \tau},\]
where $\beta$ is a user-defined constant allowing the source function to increase or decrease with optical depth. To prevent divergent integrals, we must have that $\beta < 1$. For $\beta > 0$, the source function increases as one goes further into the star and we expect limb darkening, while $\beta < 0$ gives limb brightening. As $\tau_0 \rightarrow 0$, we obtain optically-thin behaviour, thus this two-parameter model gives the desired modelling freedom.

For an incident angle $\theta$ such that $\mu = \cos \theta$, the emergent intensity is then given by
\[ I(\mu) \propto \int_0^{\tau_0} e^{\beta \tau} e^{-\tau/\mu}\frac{d\tau}{\mu},\]
where the variable of integration $\tau$ is the vertical optical depth while the optical depth along the line of sight is $\tau/\mu$. Therefore 
\[ I(\mu) \propto \frac{1 - \exp (\beta-1/\mu) \tau_0}{1 - \beta \mu} .\]
In the limit $\tau_0 \rightarrow \infty$ we have
\[ I(\mu) \propto \frac{1}{1 - \beta \mu} ,\]
which for small $\beta$ gives $I(\mu) \propto 1 + \beta \mu$, giving limb-darkening or brightening as expected. In the optically thin limit,
$\tau_0 \rightarrow 0$, 
\[ I(\mu) \propto \frac{\tau_0}{\mu}.\]
The $\mu$ divides out with the $\mu$ factor from Lambert's law, and we find that each unit area contributes equally to all directions, as long as it remains visible. This enhances the star's limb compared to the optically thick case. At quadrature for example, this will enhance emission from the sub-stellar point, leading to a low semi-amplitude. We believe that this is why lines such as MgII which have light-curves close to the optically-thin case have the lowest semi-amplitudes.

\subsection{Selecting $\tau_0$ and $\beta$}

The only information we have at hand for selecting values of $\tau_0$ and $\beta$ are the light curves of the line fluxes. In principle the line widths should help too, but more sophisticated models than ours are needed to understand these. The values we obtained are listed in Table~\ref{irradprof}. Once defined, it was a simple matter to adjust $K_\mathrm{sec}$ to match the observed values. The reduced scatter in the corrected $K$ values compared to the directly measured ones visible in Figure~\ref{kcorr} suggests that our model, although crude, has some elements of truth underlying it. Nevertheless, it rests upon several untested assumptions and a future goal should be to model the irradiated lines in a physically-consistent manner along the lines of \citet{Barman04} because then we will be able to be more certain of the $K$-correction needed in NN Ser.

\label{lastpage}

\end{document}